\documentclass[%
reprint,
superscriptaddress,
amsmath,amssymb,
aps,
pra,
]{revtex4-2}
\usepackage{multirow}
\usepackage{booktabs}
\usepackage{floatflt}
\usepackage{ragged2e}
\renewcommand{\justify}{\leftskip=0pt \rightskip=0pt plus 0cm}
\usepackage{floatrow}
\usepackage{graphicx}
\usepackage{subfigure}
\usepackage{filecontents}
\usepackage{pgfplots}
\usepackage{amsmath}
\usepackage{amssymb}
\usepackage{tikz}
\usepackage{dcolumn}
\usepackage{bm}
\usepackage{blindtext}
\usepackage{listings}
\usepackage{physics}
\usepackage{stmaryrd}
\usepackage[colorlinks,linkcolor=blue,anchorcolor=blue,citecolor=blue]{hyperref}
\makeatletter
\newif\if@restonecol
\makeatother

\usepackage[ruled,linesnumbered,vlined]{algorithm2e}
\usepackage{algpseudocode}

\begin{document}
\title{Concatenation of the Gottesman-Kitaev-Preskill code with the XZZX surface code}

\author{Jiaxuan Zhang}
\affiliation{Key Laboratory of Quantum Information, Chinese Academy of Sciences, School of Physics, University of Science and Technology of China, Hefei, Anhui, 230026, P. R. China}
\affiliation{CAS Center For Excellence in Quantum Information and Quantum Physics, University of Science and Technology of China, Hefei, Anhui, 230026, P. R. China}
\affiliation{Hefei National Laboratory, University of Science and Technology of China, Hefei 230088, P. R. China}

\author{Yu-Chun Wu}
\email{wuyuchun@ustc.edu.cn}
\affiliation{Key Laboratory of Quantum Information, Chinese Academy of Sciences, School of Physics, University of Science and Technology of China, Hefei, Anhui, 230026, P. R. China}
\affiliation{CAS Center For Excellence in Quantum Information and Quantum Physics, University of Science and Technology of China, Hefei, Anhui, 230026, P. R. China}
\affiliation{Hefei National Laboratory, University of Science and Technology of China, Hefei 230088, P. R. China}
\affiliation{Institute of Artificial Intelligence, Hefei Comprehensive National Science Center, Hefei, Anhui, 230088, P. R. China}

\author{Guo-Ping Guo}
\affiliation{Key Laboratory of Quantum Information, Chinese Academy of Sciences, School of Physics, University of Science and Technology of China, Hefei, Anhui, 230026, P. R. China}
\affiliation{CAS Center For Excellence in Quantum Information and Quantum Physics, University of Science and Technology of China, Hefei, Anhui, 230026, P. R. China}
\affiliation{Hefei National Laboratory, University of Science and Technology of China, Hefei 230088, P. R. China}
\affiliation{Institute of Artificial Intelligence, Hefei Comprehensive National Science Center, Hefei, Anhui, 230088, P. R. China}
\affiliation{Origin Quantum Computing Hefei, Anhui 230026, P. R. China}
\date{\today}
\begin{abstract}
Bosonic codes provide an alternative option for quantum error correction. An important category of bosonic codes called the Gottesman-Kitaev-Preskill (GKP) code has aroused much interest recently. Theoretically, the error correction ability of GKP code is limited since it can only correct small shift errors in position and momentum quadratures. A natural approach to promote the GKP error correction for large-scale, fault-tolerant quantum computation is concatenating encoded GKP states with a stabilizer code. The performance of the XZZX surface-GKP code, i.e., the single-mode GKP code concatenated with the XZZX surface code is investigated in this paper under two different noise models. Firstly, in the code-capacity noise model, the asymmetric rectangular GKP code with parameter $\lambda$ is introduced. Using the minimum weight perfect matching decoder combined with the continuous-variable GKP information, the optimal threshold of the XZZX-surface GKP code reaches $\sigma\approx0.67$ when $\lambda=2.1$, compared with the threshold $\sigma\approx0.60$ of the standard surface-GKP code.
Secondly, we analyze the shift errors of two-qubit gates in the actual implementation and build the full circuit-level noise model. By setting the appropriate bias parameters, the logical error rate is reduced by several times in some cases. These results indicate the XZZX surface-GKP codes are more suitable for asymmetric concatenation under the general noise models. We also estimate the overhead of the XZZX-surface GKP code which uses about 291 GKP states with the noise parameter 18.5 dB ($\kappa/g \approx 0.71\%$) to encode a logical qubit with the error rate $2.53\times10^{-7}$, compared with the qubit-based surface code using 3041 qubits to achieve almost the same logical error rate.
\end{abstract}
\maketitle
\section{Introduction}\label{s1}
Quantum computation is promised to offer speedups over the best-known classical algorithms for solving certain types of problems \cite{nielsen2002quantum,shor1999polynomial,feynman1985quantum,hallgren2007polynomial,freedman2002simulation,lloyd2010quantum}. To realize large-scale practical quantum computation, the quantum error correction is a crucial problem since physical quantum states are too fragile to be preserved in the uncontrolled environment \cite{terhal2015quantum,gottesman1997stabilizer,preskill1998reliable,steane1996multiple,bennett1996mixed}. 

The fundamental idea of the quantum error correction is introducing redundancy by encoding logical qubits in a high-dimensional Hilbert space. Unlike the two-level qubit-based systems, continuous-variable quantum systems provide attractive alternative for the quantum error correction \cite{braunstein2005quantum,weedbrook2012gaussian,adesso2014continuous,ferraro2005gaussian,cerf2007quantum}. The bosonic codes encode the quantum information in bosonic modes which provide infinite-dimensional Hilbert space \cite{albert2018performance,cai2021bosonic}. With the rapid developments in quantum hardware and control technology, bosonic codes have shown their unprecedented potential in quantum error correction in many different experiments  \cite{ofek2016extending,hu2019quantum,campagne2020quantum,de2022error,rosenblum2018fault}. In the last few years, bosonic quantum error correction has attracted a lot of interest, since it is demonstrated to reach the break-even point \cite{ofek2016extending,sivak2022real,ni2023beating}, i.e., the lifetime of a logical qubit is enhanced to exceed that of any individual components composing the experimental system. The representative bosonic codes based on a single bosonic mode include the cat code \cite{guillaud2022quantum,leghtas2013hardware,ofek2016extending}, the binomial code \cite{michael2016new,albert2018performance,hu2019quantum} and the Gottesman-Kitaev-Preskill (GKP) code \cite{gottesman2001encoding,grimsmo2021quantum}.

The GKP code was proposed by Gottesman, Kitaev, and Preskill in 2001, encoding the qubit into a harmonic oscillator \cite{gottesman2001encoding}. It was considered to be impractical for a long time, but now arouses extensive attention because of the recent experimental realizations \cite{campagne2020quantum,de2022error,sivak2022real}. Theoretically, the GKP-qubit encoding is close to the optimal encoding for the quantum capacity of Gaussian thermal loss channels with average photon-number constraint \cite{noh2018quantum}. Nevertheless, the protection of quantum information provided by the GKP code is limited. The GKP code is helpless for the shift error beyond a certain boundary, in which case the logical error may be produced.

To overcome the logical errors of the GKP state, it is natural to introduce a high-level stabilizer code \cite{wang2019quantum}. As the main representatives of the two-dimensional topological stabilizer codes, surface codes \cite{fowler2009high,fowler2012surface} and color codes \cite{bombin2006topological,fowler2011two} concatenated with GKP code have been considered in many previous works \cite{vuillot2019quantum,noh2020fault,noh2022low,zhang2021quantum}, which discuss the performance of the concatenation codes in different noise models. 

Ref.$\,$\cite{bonilla2021xzzx} proposes a variant of the surface code -- the XZZX surface code which shows its high threshold and low overhead under biased noise. It should be noted that even if the bias is equal to $1$, the threshold of the XZZX surface code using the minimum weight perfect matching (MWPM) decoder \cite{edmonds1965paths,higgott2021pymatching} is still slightly higher than the conventional surface code. Therefore, it is natural to expect a better performance of the concatenation of the GKP code with the XZZX surface code  than the conventional surface code.

The goal of this paper is to study the concatenation of the GKP code with the XZZX surface code (the XZZX surface-GKP code). The performance of the XZZX surface-GKP code is investigated under two noise models. Concretely, in the first noise model, the code-capacity noise model \cite{stephens2014fault}, all the components except data GKP qubits are noiseless. For utilizing the advantage of the XZZX surface code in handling biased noise, the asymmetric rectangular GKP code \cite{hanggli2020enhanced} with parameter $\lambda$ is introduced to create the bias artificially. Using the MWPM decoder combined with the continuous-variable GKP information, the optimal code-capacity threshold of the XZZX-surface GKP code reaches $\sigma\approx 0.67$ when $\lambda=2.1$, exceeding the previous result of the standard GKP-surface code with the threshold $\sigma\approx 0.60$ \cite{vuillot2019quantum}. Another work  about the asymmetric GKP concatenation studies the rectangular GKP code concatenated with the conventional surface code \cite{hanggli2020enhanced}, where they use the BSV decoder \cite{bravyi2014efficient} without the GKP continuous-variable information and improve the threshold from $\sigma\approx0.54$ to $\sigma\approx0.58$. Compared with their result, the XZZX-surface GKP code is a more promising candidate for asymmetric GKP concatenation code. 

The second noise model called the full circuit-level noise model is more realistic, where the noises in GKP state preparations, homodyne measurements, two-qubit gates and idle operations are taken into account. This error model is built on the specific derivation of the Gaussian shift errors after CNOT, CZ gates, and balanced beam-splitter operations. The maximum likelihood (ML) decoding strategy in Ref.$\,$\cite{noh2022low} is adapted to the full circuit-level noise model. The threshold of the XZZX-surface GKP code under this noise model reaches 16.1 dB ($\kappa/g \approx 1.23\%$), which is superior to the previous result 18.6 dB ($\kappa/g \approx 0.69\%$) of the surface GKP code under the same error model \cite{noh2020fault}. Meanwhile, by setting the appropriate bias parameters, the logical error rate is reduced by several times compared with the square GKP concatenation code. Finally, we estimate the overhead of the XZZX-surface GKP code to achieve a logical error rate that is low enough. For example, if the noise parameter reaches 18.5 dB ($\kappa/g \approx 0.71\%$), one needs about 291 GKP states to encode a logical qubit with the error rate $2.53 \times 10^{-7}$, which is much fewer than the qubit-based surface code.

The rest of the paper is organized as follows. Section \ref{s2} starts with some basic aspects of the conventional surface code and the XZZX surface code. Then we discuss some decoding algorithms, especially the MWPM algorithm which will be applied in the decoding of the whole following paper. Section \ref{s3} introduces the asymmetric rectangular GKP code and the construction and error correction circuits of the XZZX surface-GKP code. At the end of this section, the code capacity threshold of the XZZX surface-GKP code with the designed bias is obtained by numerical simulations.  In Section \ref{s4} we discuss the full circuit-level noise model and the ML decoding strategy of two-qubit gates in detail. Section \ref{s4} also presents the numerical results and estimates the overhead of the XZZX surface-GKP code. Lastly, the conclusion of this paper and the outlook for future work are described in Section \ref{s5}.

\begin{figure*}
\centering
\subfigure[]{
\label{fig1a}
\includegraphics[width=5.1cm]{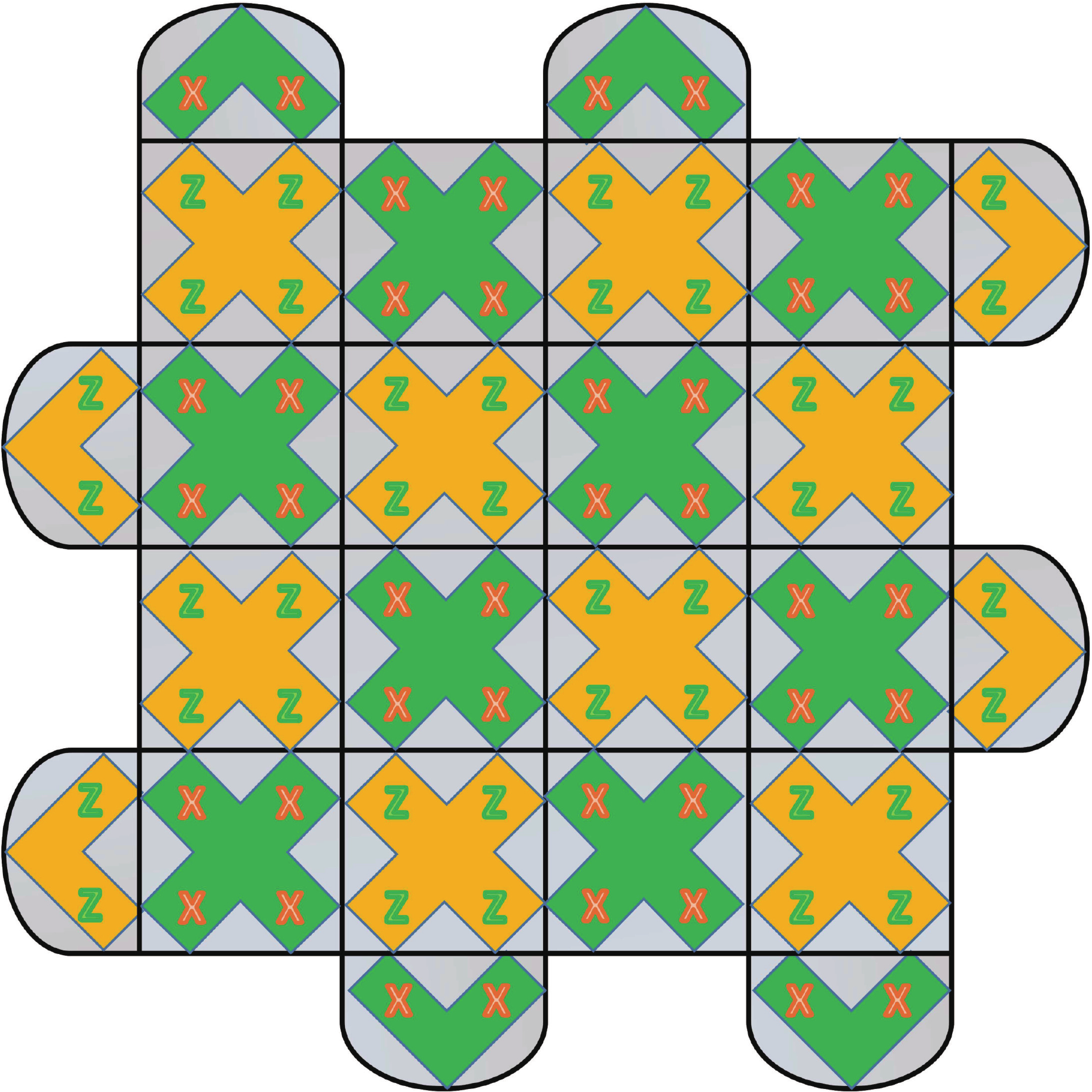}}
\hspace{0.1in}
\subfigure[]{
\label{fig1b}
\includegraphics[width=5.1cm]{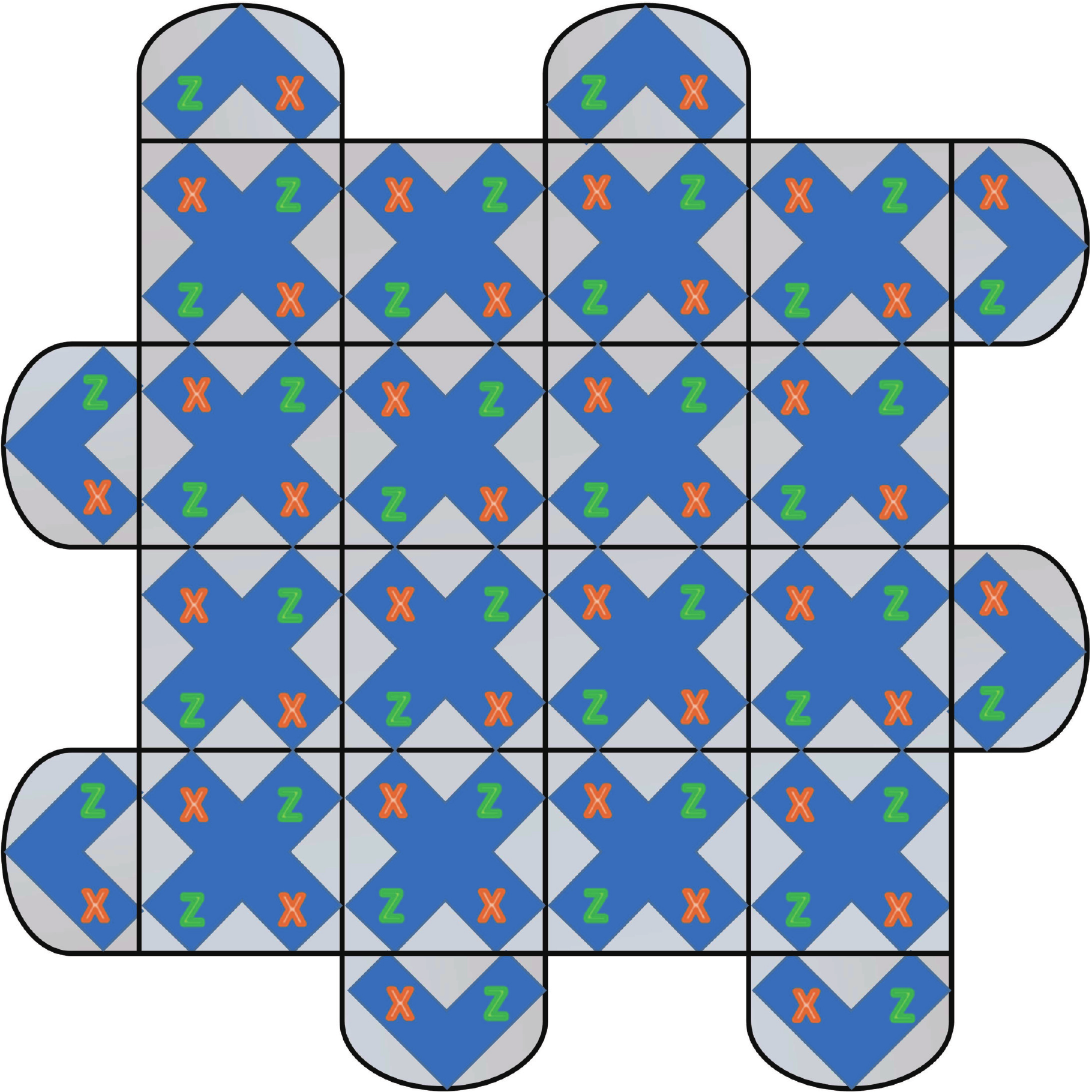}} 
\hspace{0.1in}
\subfigure[]{
\label{fig1c}
\includegraphics[width=5.1cm]{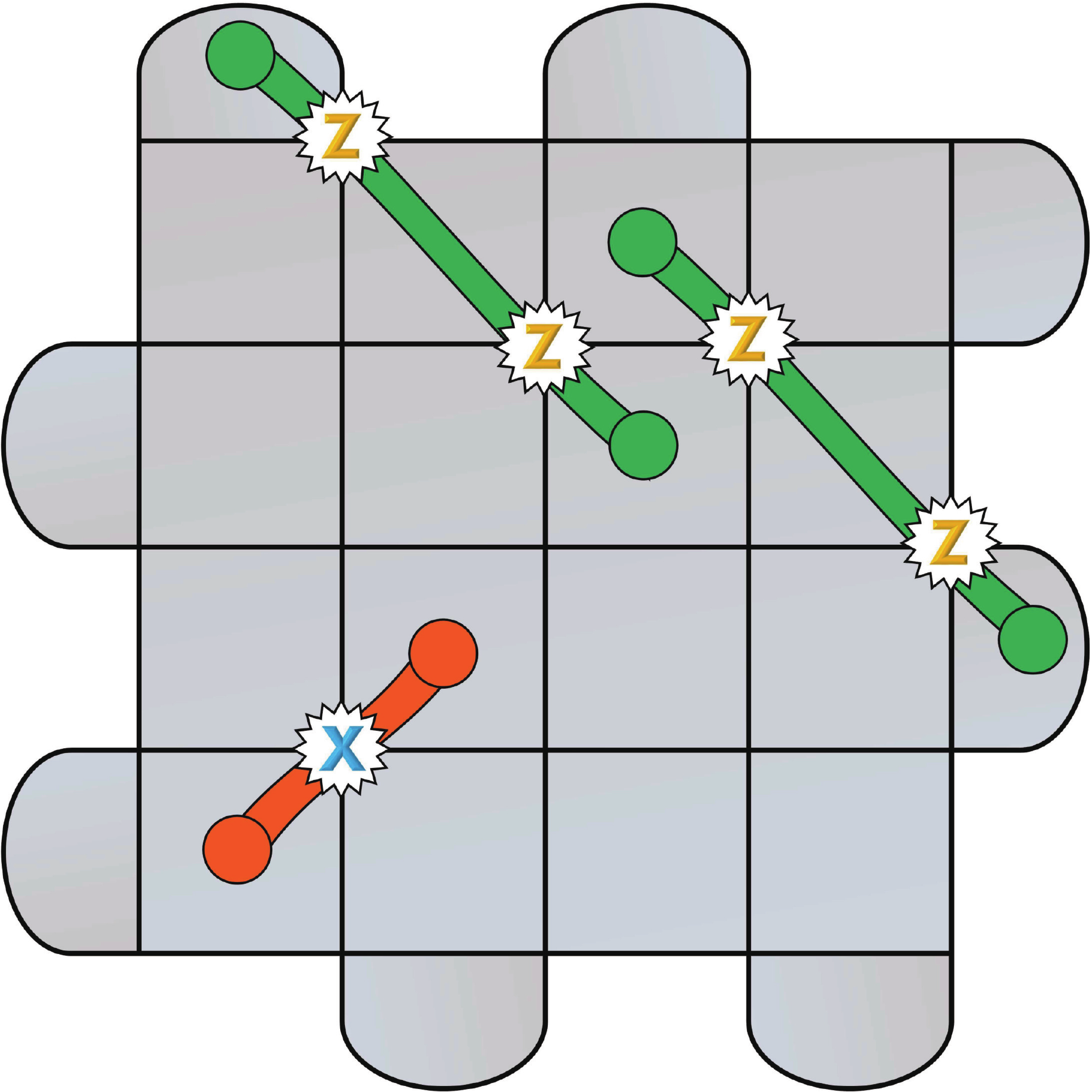}}  
\caption{\justify (Color online) Two types of rotated surface codes and error strings. (a) The conventional surface code.  Data qubits lie on the vertices of each square face, and each face indicates a stabilizer, which is the X (or Z) tensor product of the adjacent data qubits. The $X$ type stabilizers and the $Z$ type stabilizers are green and yellow respectively. (b) The XZZX surface code. The XZZX surface code only has XZZX-type stabilizers which is blue in the figure. (c) Error strings of the XZZX surface code. The labeled stabilizers (bold dots on the faces) of the XZZX surface code caused by only $X$ or $Z$ errors can be connected with disjoint strings. Each string can be viewed as a repetition code.}
\end{figure*}

\section{The XZZX surface code}\label{s2}
\subsection{the conventional surface code}
The surface code is a kind of topological stabilizer codes. Let $P_n$ denote the $n$-fold Pauli group, the elements in which are $n$-fold tensor products of the single qubit Pauli operators $\{I,X,Y,Z\}$ with the phase $\pm 1$ or $\pm i$. Given a quantum state $\ket{\psi}$, an operator $S\in P_n$ is called the stabilizer of $\ket{\psi}$ if $S \ket{\psi}=+1\ket{\psi}$. 

In the stabilizer formalism \cite{poulin2005stabilizer}, a logical state of the surface code is described by an abelian subgroup called the stabilizer group in which all the operators are the stabilizers of the quantum state. Note that every two stabilizer operators in a stabilizer group commute and the identity operator $I$ is always in the stabilizer group. The stabilizer group of a logical state of the surface code is generated by two types of operators -- $X$ type stabilizers and $Z$ type stabilizers, which are the XXXX tensor product and ZZZZ tensor product of four data qubits in each square face (see Fig.$\,$\ref{fig1a}). 

Let us focus on the rotated surface code which uses fewer data qubits to encode a logical qubit compared with the planar surface code \cite{bombin2007optimal}. The rotated surface code can be described intuitively in a two-dimensional regular lattice, as shown in Fig.$\,$\ref{fig1a} The data qubits which are encoded to store quantum information lie on the vertices, and each stabilizer operator corresponds to a colored face. Concretely, the form of $X$ (or $Z$) type stabilizers is $S_X=\prod_{i\in \partial f}X_i$ (or $S_Z=\prod_{i\in \partial f}Z_i$), where $\{i\}$ are the labels of the vertices incident on face $f$. Note that the boundary stabilizers have the weight of 2 and the other stabilizers have the weight of 4 since the boundary face only has two vertices.

The surface code reduces the logical error rate efficiently by increasing the code distance $d$. The code distance $d_x$ (or $d_z$) is the Pauli weight of the logical Pauli operator $X_L$ (or $Z_L$), that is, the $X$ (or $Z$) tensor product of the data qubit in the string between two horizontal (or vertical) boundaries. If the error rate of each data qubit $p$ is small enough, the logical error rate $P_L$ can be approximated well by the empirical formula \cite{fowler2012surface}
\begin{equation}\label{e1}
    P_L\cong c \,(p/p_{th})^{\frac{d+1}{2}},
\end{equation}
where $p_{th}$ is an important parameter in the quantum error correction theory called the threshold. Apparently, if the data qubit error rate is less than the threshold, the logical error rate tends to be infinitely small with the increase of the code distance.

\subsection{the XZZX surface code}
In many original works of surface code threshold, the noise model is the single-qubit Pauli noise channel where the data qubit suffers the $X$, $Y$, and $Z$ error with equal probabilities. However, in many experimental architectures \cite{aliferis2009fault,shulman2012demonstration,nigg2014quantum}, the realistic noise is biased towards dephasing, i.e., the $Z$ error rate is much higher than the $X$ and $Y$ error's. Usually, the noise bias is defined as $\eta=p_z/(p_x+p_y)$, where $p_x$, $p_y$, $p_z$ are the Pauli $X$, $Y$, $Z$ error rate of the qubits respectively.

In order to better adapt to the realistic noise model, some well-designed surface codes are developed, one of which is the XZZX surface code \cite{tuckett2018ultrahigh,tuckett2020fault,bonilla2021xzzx}. It has been shown numerically that the XZZX surface code has a high threshold under the biased noise model, which exceeds the hashing bound \cite{bennett1996mixed}.

The XZZX surface code can be defined in the same 2D lattice as the conventional surface code. Without regard to the boundary, there is only one type stabilizer generators of the XZZX surface -- the tensor product of Pauli operator XZZX. Note that the XZZX surface code can be converted to the conventional surface code by locally acting Hadamard operators on alternate data qubits.

Suppose that some Pauli errors occur in the surface code, the stabilizer generators are labeled if the generators anti-commute with the Pauli errors. A good characteristic of the XZZX surface code is that the labeled stabilizers caused by only $Z$ (or $X$) errors can be connected with disjoint strings, as shown in Fig.$\,$\ref{fig1c}. Under the $Z$ error-dominated noise, these parallel strings can be viewed as in the independent repetition codes \cite{chen2021exponential} if we ignores the $X$ or $Y$ errors. It is well known that the repetition code has an ideal threshold $p_{th}=0.5$, which is much higher than the conventional surface code's $0.1$. Therefore, the threshold of the XZZX surface code is superior to the conventional surface code under a bias dephasing noise model.

Fig.$\,$\ref{fig1b} shows a $d=5$ XZZX surface code where the logical Z and X operator have the equal length. However, because of the biased noise, the logical Z operator is longer than the logical X operator in the practical XZZX surface code. In other words, the array of the data qubits is typically a rectangle rather than a square, which provides the approach for XZZX surface codes to achieve a target logical failure rate with low overhead.

\subsection{decoding}
Turn our attention to the decoding problem of the surface code. Assuming $\mathcal{S}$ generated by $\{S_k\}$ is the stabilizer group of a quantum code $C$, a $n$-qubit Pauli operator $e\in P_n$ is a detectable error if there exists at least one stabilizer $S_i\in \{S_k\}$ such that $S_i$ anticommutes $e$. The outcomes of measuring all these stabilizer generators form the syndromes of error $e$, denoted by $s_e$. Suppose an error $e$ occurs in some data qubits with the syndrome $s_e$, a decoding algorithm outputs a recover operator $R$ according to the syndrome, such that $Re$ commutes with all the stabilizers in $\mathcal{S}$.

However, the decoding algorithm does not always output an expected $R$. The error $e$ is corrected successfully if $Re\in\mathcal{S}$, otherwise the decoding process produces a logical error that cannot be detected by the stabilizer measurements. Generally, the decoding algorithm is hoped to have the logical error rate as low as possible.

In fact, for a given syndrome $s_e$, the recover operator $R$ may come from one of the following four sets:
\begin{equation}
e\mathcal{S},\quad
eX_L\mathcal{S},\quad
eZ_L\mathcal{S},\quad
eX_LZ_L\mathcal{S}.
\end{equation}
These four sets are the cosets of $\mathcal{S}$ and here we assume the quantum error correction code only encodes one logical qubit. The recover operators in the same coset are equivalent when we recover an error state. Selecting a recover operator from the most likely coset will reduce the logical error rate, which is exactly what the maximum-likelihood
decoder (MLD) does \cite{dennis2002topological}.

The MLD is optimal from the point of getting a high success probability of decoding. However, the exact MLD under a general noise model costs a lot of time to execute. Compared with the MLD, the MWPM decoder is more widely-used, because of its efficiency.

The MWPM decoder turns the decoding problem into matching the labeled stabilizers in pairs with the lowest link weight. Firstly it transforms the surface code lattice into a decoding graph where the vertices denote the stabilizers and each data qubit lies on the edge \cite{wang2011surface}. Consider a path $E$ on the decoding graph where $E$ corresponds to an error (say $e$) whose labeled stabilizers are the starting and ending vertices of $E$. The probability of occurring the error $e$ is
\begin{equation}\label{e3}
\begin{aligned}
p_E&=\prod_{i\not \in E}(1-p_{i})
\times\prod_{i \in E}p_{i}\\
&=P_0\prod_{i \in E}(\frac{p_{i}}{1-p_{i}}),
\end{aligned}
\end{equation}
where $i$ is the label of the edge, $p_{i}$ is the error rate of the data qubits adjacent to $E$, and $P_0=\prod_{i}(1-p_{i})$ is a constant irrelevant to $E$. Therefore, if one sets the weight of the edge $i$ as $w_i=-\log\frac{p_{i}}{1-p_{i}}$, $-\log p_E$ is exactly the sum of the weights of the edge in path $E$ (up to a constant):
\begin{equation}\label{e4}
\begin{aligned}
-\log p_E = \sum_{i\in E}w_i-\log P_0.
\end{aligned}
\end{equation}

Given the weight of each possible path, the minimum weight perfect matching algorithm can be applied to connect the labeled stabilizers in pairs such that the sum of weights is minimal. Then the recover operators are applied along the connecting path. As mentioned, executing the MWPM decoder is more efficient since the time cost of the minimum weight perfect matching algorithm is polynomial. Although the MWPM decoder is not optimal, it still performs well enough in many usual noise models. More importantly, the MWPM decoder can naturally adapt to the surface code decoding adding the GKP continuous-variable information by modifying the matching weights, which is discussed in the following sections.

\section{the XZZX surface-GKP code}\label{s3}
In this section, we introduce the main topic of this paper -- the XZZX surface-GKP code. Specifically, this section describes the concatenation of the rectangular GKP code with the XZZX surface code in several aspects including basic concepts, construction, and error correction process. At the end of this section, we investigate the performance of the XZZX surface-GKP code in the code-capacity noise model and show the numerical results.
\subsection{the rectangular GKP code}
The GKP error correction code is a kind of bosonic code that has attracted much attention recently. Many previous works focus on the GKP code on the square lattice (square GKP code). The square GKP code protects against small shift error in position and momentum quadratures with equal logical $\bar{X}$ or $\bar{Z}$ error rates. As a more general case, here we review the basic aspects of the rectangular GKP code. 

For a harmonic oscillator,  the position and momentum operators are defined as:
\begin{equation}\label{e5}
\hat{q}=\frac{1}{\sqrt{2}}(\hat{a}+\hat{a}^\dag),\quad
\hat{p}=-\frac{i}{\sqrt{2}}(\hat{a}-\hat{a}^\dag),
\end{equation}
where $\hat{a}^\dag$ and $\hat{a}$ are annihilation and creation operators satisfying $[\hat{a}^\dag,\hat{a}]=1$. The codespace of rectangular GKP code is stabilized by two commuting stabilizer operators 
\begin{equation}\label{e6}
\hat{S}_{p,\lambda}=e^{-i2(\sqrt{\pi}\lambda)\hat{p}},\quad
\hat{S}_{q,\lambda}= e^{i2(\sqrt{\pi}/\lambda)\hat{q}},
\end{equation}
where $\lambda$ is the parameter of asymmetry. The logical states are defined as:
\begin{equation}\label{e7}
\begin{aligned}
&|\bar{0}_\lambda\rangle \propto 
\sum_{n\in \mathbb{Z}} |{q}=2n\lambda\sqrt{\pi} \rangle,\\
&|\bar{1}_\lambda\rangle \propto 
\sum_{n\in \mathbb{Z}} |{q}=(2n+1)\lambda\sqrt{\pi} \rangle .
\end{aligned}
\end{equation}
Correspondingly,  the logical Pauli operators are:
\begin{equation}\label{e8}
\bar{X}_\lambda=e^{-i(\sqrt{\pi}\lambda)\hat{p}},\quad
\bar{Z}_\lambda= e^{i(\sqrt{\pi}/\lambda)\hat{q}}.
\end{equation}

The Clifford gates of the GKP code can be performed by interactions that are at most quadratic in the creation and annihilation operators. In particular, the rescaled $\mathrm{CNOT}_{\beta}$ gate with parameter $\beta$ has the following form:
\begin{equation}\label{e9}
\begin{aligned}
\mathrm{CNOT}_{\beta} =e^{-i\hat{q}_j\hat{p}_k/\beta},
\end{aligned}
\end{equation}
where $j$, $k$ are the control qubit and target qubit, respectively. The $\mathrm{CNOT}_{\beta}$ gate holds the following relationships with $\hat{q}$ and $\hat{p}$ operators in the Heisenberg representation \cite{gottesman1998heisenberg}:
\begin{equation}\label{e10}
\begin{aligned}
&(\mathrm{CNOT}_{\beta})^{-1}\hat{q}_j\mathrm{CNOT}_{\beta}=\hat{q}_j,\\
&(\mathrm{CNOT}_{\beta})^{-1}\hat{q}_k\mathrm{CNOT}_{\beta}=\hat{q}_k+\hat{q}_j/\beta,\\
&(\mathrm{CNOT}_{\beta})^{-1}\hat{p}_j\mathrm{CNOT}_{\beta}=\hat{p}_j-p_k/\beta,\\
&(\mathrm{CNOT}_{\beta})^{-1}\hat{p}_k\mathrm{CNOT}_{\beta}=\hat{p}_k.
\end{aligned}
\end{equation}
Likewise, the rescaled $\mathrm{CZ}_{\beta}$ gate is 
\begin{equation}\label{e11}
\begin{aligned}
\mathrm{CZ}_{\beta} =e^{-i\hat{q}_j\hat{q}_k/\beta}.
\end{aligned}
\end{equation}
One can easily find how the $\mathrm{CZ}_{\beta}$ transforms $\hat{q}$ and $\hat{p}$ in the Heisenberg representation by exchanging $\hat{q}_k$ and $\hat{p}_k$ in Eq.$\,$(\ref{e10}), since $\mathrm{CZ}_{\beta}=H_k^{-1}\mathrm{CNOT}_{\beta}H_k$ where $H_k=e^{i\pi\hat{a}^\dag\hat{a}}$ is the Hadamard gate of qubit $k$.

Another important operation in the error correction of the GKP code is the beam-splitter 
\begin{equation}\label{e12}
\begin{aligned}
B_{\theta} =e^{-i\theta(\hat{q}_j\hat{p}_k-\hat{p}_j\hat{q}_k)},
\end{aligned}
\end{equation}
which transforms $\hat{q}$ and $\hat{p}$ as follows:
\begin{equation}\label{e13}
\begin{aligned}
&(B_{\theta})^{-1}\hat{q}_jB_{\theta}=\cos{\theta} \hat{q}_j-\sin{\theta} \hat{q}_k,\\
&(B_{\theta})^{-1}\hat{q}_kB_{\theta}=\cos{\theta} \hat{q}_k+\sin{\theta} \hat{q}_j,\\
&(B_{\theta})^{-1}\hat{p}_jB_{\theta}=\cos{\theta} \hat{p}_j-\sin{\theta} \hat{p}_k,\\
&(B_{\theta})^{-1}\hat{p}_kB_{\theta}=\cos{\theta} \hat{p}_k+\sin{\theta} \hat{p}_j.
\end{aligned}
\end{equation}
We call this operation balanced beam-splitter if $\theta=\pi/4$.

Generally, the noise channel of the GKP code is considered as a Gaussian shift error channel $\mathcal{N}$ \cite{wang2019quantum}:
\begin{equation}\label{e14}
 \mathcal{N}(\rho)\equiv \iint  P_\sigma (u)  P_\sigma (v) e^{-iu\hat{p}} e^{iv\hat{q}} \rho e^{-iv\hat{q}} e^{iu\hat{p}}du dv,
\end{equation}
where $P_\sigma (x)=\frac{1}{\sqrt{2\pi \sigma^2}}e^{-\frac{x^2}{2\sigma^2}}$ is Gaussian distribution function with variance $\sigma^2$. Such an error channel is the result of Pauli twirling approximation \cite{katabarwa2017dynamical}. An logical $\bar{X}$ (or $\bar{Z}$) error occurs when the Gaussian shift error $|u\mod{2\sqrt{\pi}}\lambda|>\frac{\sqrt{\pi}}{2}\lambda$ (or $|v\mod{2\sqrt{\pi}}/\lambda|>\frac{\sqrt{\pi}}{2\lambda}$) in the $\hat{q}$ (or $\hat{p}$) quadrature. In this paper, the mod function $a\mod{b}$ has the range $[-\frac{b}{2},\frac{b}{2})$. 

The logical $X$ and $Z$ error probabilities are  
\begin{equation}\label{e15}
\begin{aligned}
&p_{\bar{X}}=1-\sum_{k\in\mathbb{Z}}\int_{2k\sqrt{\pi}\lambda-\frac{\sqrt{\pi}\lambda}{2}}^{2k\sqrt{\pi}\lambda+\frac{\sqrt{\pi}\lambda}{2}} P_\sigma (u) du,\\
&p_{\bar{Z}}=1-\sum_{k\in\mathbb{Z}}\int_{2k\sqrt{\pi}/\lambda-\frac{\sqrt{\pi}}{2\lambda}}^{2k\sqrt{\pi}/\lambda+\frac{\sqrt{\pi}}{2\lambda}} P_\sigma (v) dv.
\end{aligned}
\end{equation}
When the error shift goes beyond the blue area in Fig.$\,$\ref{fig2a}, the error correction procedure will introduce a logical $X$ or $Z$ error. Intuitively, the logical error probability is the probability of the Gaussian shift error out of the correctable range in Fig.$\,$\ref{fig2a}.  

The GKP code provides extra continuous-variable information for surface code decoding. The error rates conditioned on the measurement results $q_m$ and $p_m$ are \cite{wang2019quantum}:
\begin{equation}\label{e17}
\begin{aligned}
&p({\bar{X}}|q_m)=1-\frac{\sum_{k\in\mathbb{Z}}P_\sigma (q_m-2k\sqrt{\pi}\lambda)}{\sum_{k\in\mathbb{Z}}P_\sigma (q_m-k\sqrt{\pi}\lambda)},\\
&p({\bar{Z}}|p_m)=1-\frac{\sum_{k\in\mathbb{Z}}P_\sigma (p_m-2k\sqrt{\pi}/\lambda)}{\sum_{k\in\mathbb{Z}}P_\sigma (p_m-k\sqrt{\pi}/\lambda)}.
\end{aligned}
\end{equation}
for ${\bar{X}}$ or ${\bar{Z}}$ error respectively, where $q_m$ and $p_m$ are the masurement results in the GKP error correction. These conditional error rates provide more accurate matching weights for the MWPM decoder.

Obviously, the rectangular GKP code goes back to the square GKP code if $\lambda=1$. The square GKP code protects against shift errors with same ranges in $\hat{q}$ and $\hat{p}$ quadratures. in which the probabilities of logical ${\bar{X}}$ or ${\bar{Z}}$ error are equal. The rectangular GKP code naturally owns the bias with the parameter
\begin{equation}\label{e16}
\eta=\frac{p_{\bar{Z}}-p_{\bar{Z}} p_{\bar{X}}}{p_{\bar{X}}}.
\end{equation}
For utilizing the superiority of the XZZX surface code under the biased noise, we can design the bias of the rectangular GKP code by setting $\lambda \neq 1$.
 
\subsection{XZZX surface-GKP code and its error correction}\label{s3.1}
\begin{figure}[t]
\centering
\subfigure[]{
\label{fig2a}
\includegraphics[width=8cm]{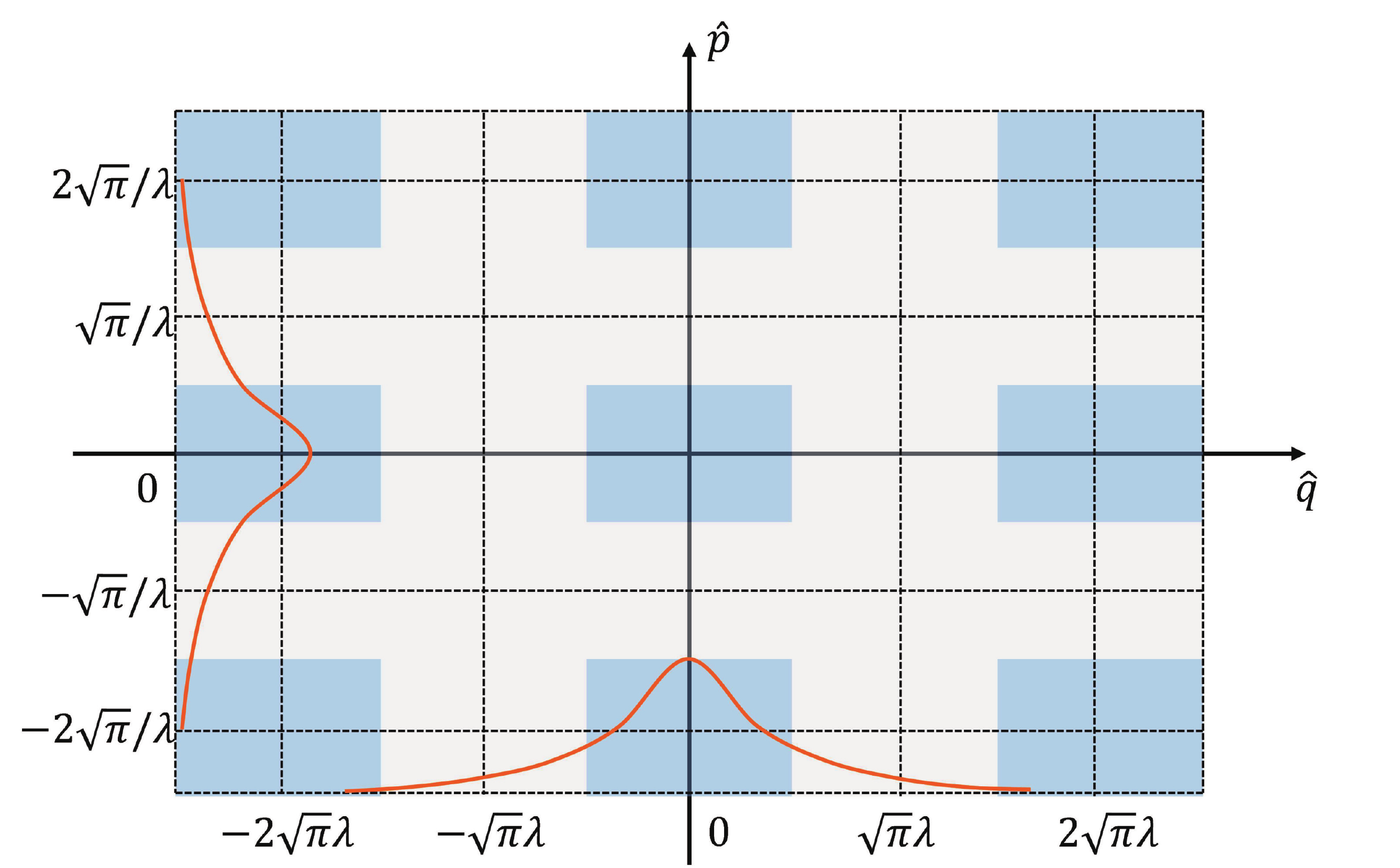}}
\vspace{0.2in}
\subfigure[]{
\label{fig2b}
\includegraphics[width=7cm]{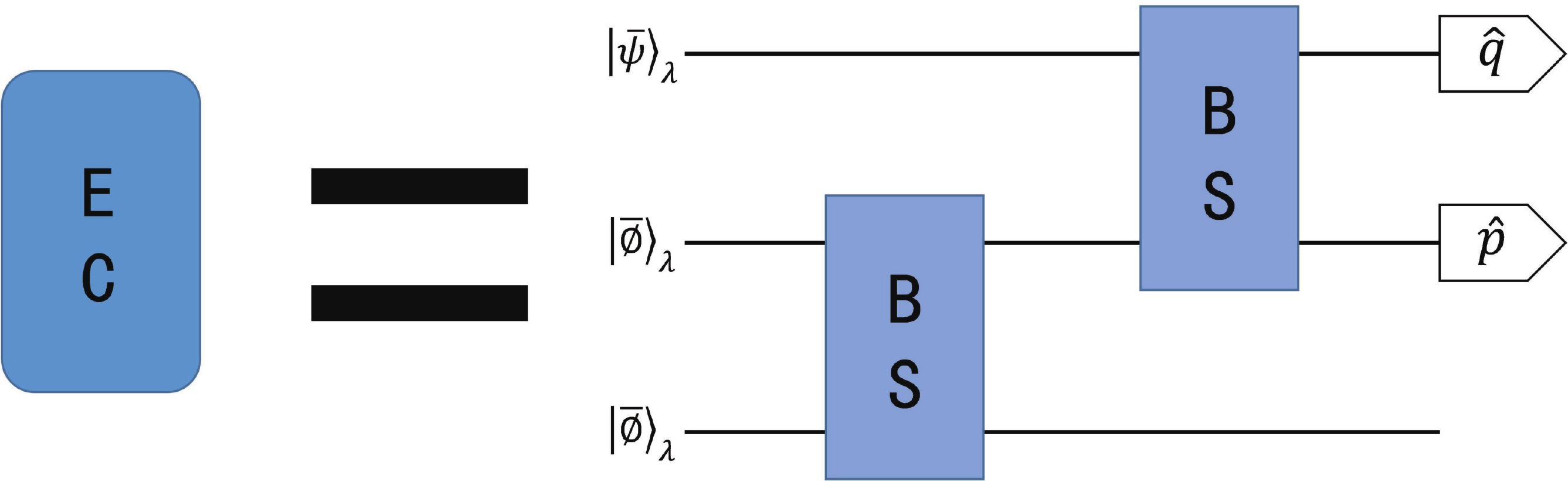}} 
\caption{\justify (Color online) The error correction process of the rectangle GKP code (a) The error correction range of rectangle GKP code in $\hat{q}$ and $\hat{p}$ quadratures. An ideal GKP error correction procedure can only correct the shift errors in the blue area. The logical error will be produced if the shift errors are beyond the blue area. (b) The quantum circuit of the teleportation-based GKP error correction (labeled as EC). Two balanced beam-splitter (labeled as BS) are used for state teleportation.}
\end{figure}

The GKP error correction code can protect against small shift errors. If the shift error is larger than $\frac{\sqrt{\pi}}{2}\lambda$ in $\hat{q}$ quadrature or $\frac{\sqrt{\pi}}{2}/\lambda$ in $\hat{p}$ quadrature, the higher level stabilizer code is necessary to deal with the GKP logical error. Now let us discuss the concatenation of the GKP code with the XZZX surface code.

\begin{figure}[t]
\centering
\subfigure[]{
\label{fig3a}
\includegraphics[width=6cm]{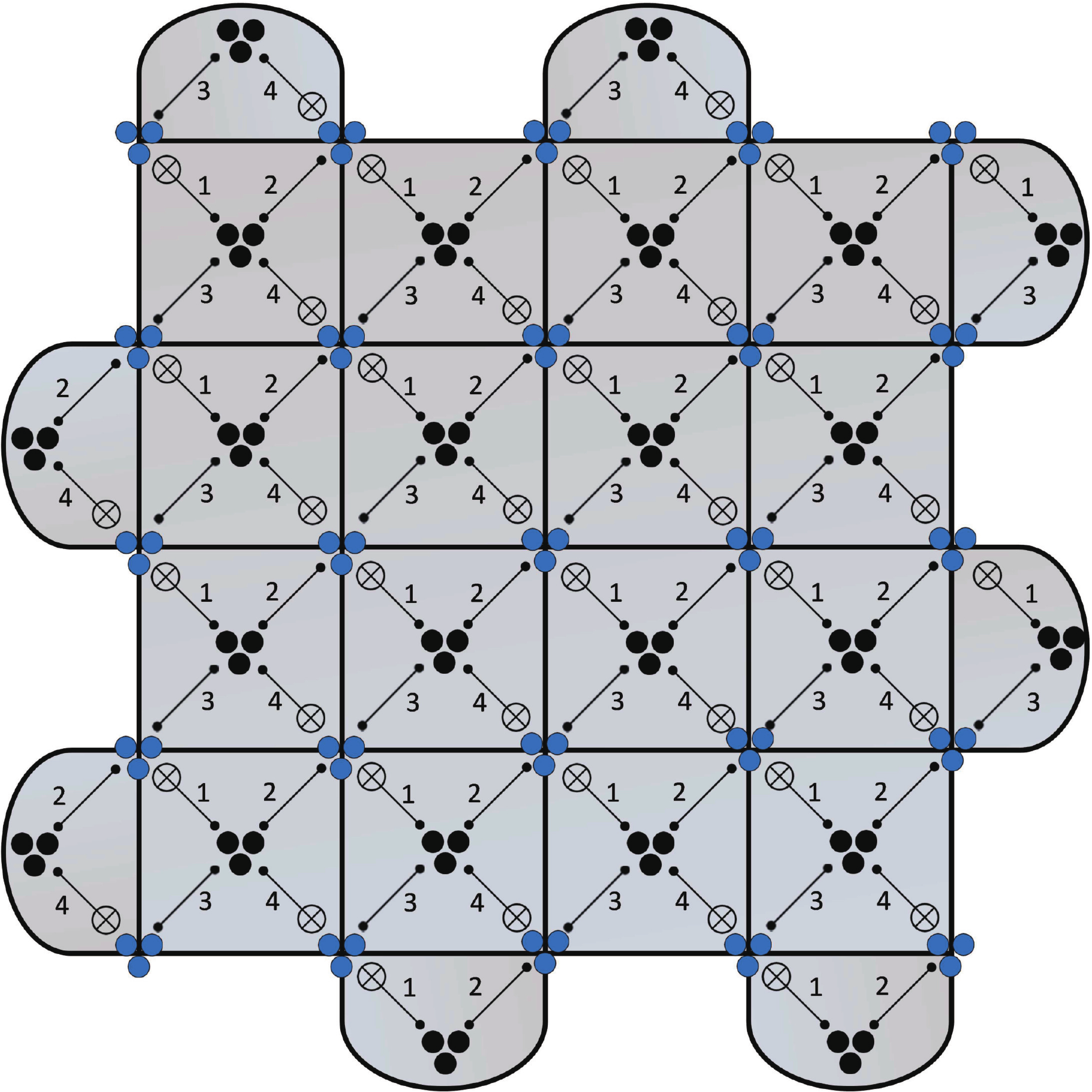}}
\vspace{0.2in}
\subfigure[]{
\label{fig3b}
\includegraphics[width=7.5cm]{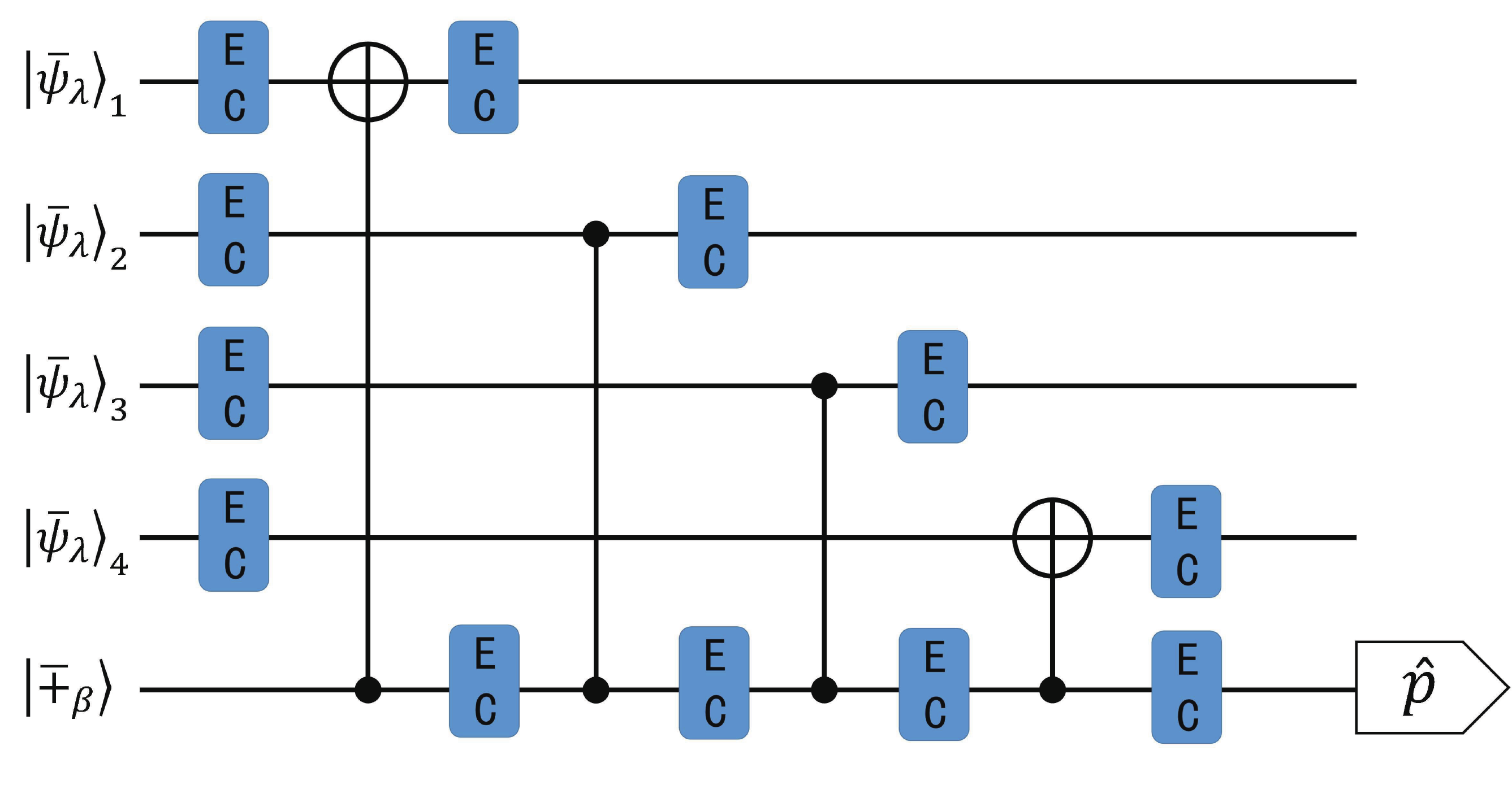}} 
\caption{\justify (Color online) The qubit layout and stabilizer measurements of the XZZX surface-GKP code. (a) The qubit layout of the XZZX surface-GKP code. Each data qubit (blue) or syndrome qubit (black) requires two extra ancilla GKP states for GKP error correction. The number near CZ or CNOT gates indicates the time order of the action. (b) The quantum circuit of the XZZX stabilizer measurement. The data and syndrome qubits are rectangle GKP states with different parameters. The CZ and CNOT gates are also rescaled. The teleportation-based error corrections are applied after CZ, CNOT gates and idle operations.}
\end{figure}

The data qubits of the XZZX surface code are replaced by encoded GKP states following the same array as Fig.$\,$\ref{fig1b}. Moreover, referring to Fig.$\,$\ref{fig3a}, two extra GKP states are placed next to each data or syndrome qubit, which is used for the GKP error correction. The GKP error correction protocol used in this paper is the teleportation-based scheme \cite{walshe2020continuous}.

The teleportation-based GKP error correction scheme utilizes the GKP Bell pair and the balanced beam-splitter operations. The GKP Bell pair is prepared from two qunaught states \cite{walshe2020continuous} 
\begin{equation}\label{e18}
\begin{aligned}
\ket{\emptyset_\lambda}\propto\sum_{n\in\mathbb{Z}}\ket{\hat{q}=n\sqrt{2\pi}\lambda}
\propto\sum_{n\in\mathbb{Z}}\ket{\hat{p}=n\sqrt{2\pi}/\lambda},
\end{aligned}
\end{equation}
acted by a balanced beam-splitter, as shown in Fig.$\,$\ref{fig2b}.  The quantum information in data GKP qubits will teleport to the second ancilla GKP qubits, where the GKP error correction is done naturally. Just like the normal quantum teleportation, the logical $\bar{X}$ or $\bar{Z}$ error may be produced, depending on the measurement results on the first two qubits. It is unnecessary to perform real-time error correction, but one can keep track of the errors in the Pauli frame \cite{knill2005quantum,chamberland2018fault} until a non-Clifford gate is applied.  

It has been shown that the teleportation-based scheme outperforms the Steane scheme where they assume the finite squeezing of the GKP states is the only noise source \cite{noh2022low}. In Appendix \ref{aa}, we present the detail of the teleportation-based scheme and give a discussion about the superiority of this scheme under a full circuit-level error model in Appendix \ref{ab}.

In the second level of the error correction, the ancilla GKP qubits called syndrome qubits are placed in each face in Fig.$\,$\ref{fig3a} for stabilizer measurements of the XZZX surface code. The syndrome qubits are coupled with nearby data qubits by $\mathrm{CNOT}$ or $\mathrm{CZ}$ gates. 

In particular, if the data qubits are the rectangular GKP state (say $|\bar{\psi}_\lambda\rangle$), the ancilla qubits and two-qubit gates need to be rescaled. Specifically, in the GKP error correction step, the GKP ancilla qubits should have the same rescaled parameter $\lambda$ as data qubits. In the XZZX stabilizer measurement step, the rescaled parameter of the syndrome qubits can be chosen arbitrarily (say $\beta$), but the $\mathrm{CNOT}$ and $\mathrm{CZ}$ gates should adjust to $\mathrm{CNOT}_{\beta/\lambda}$ and $\mathrm{CZ}_{\lambda\beta}$ gates respectively (see Appendix \ref{ac} for detail).

\begin{figure*}[t]
\centering
\subfigure[]{
\label{fig4a}
\includegraphics[width=8.2cm]{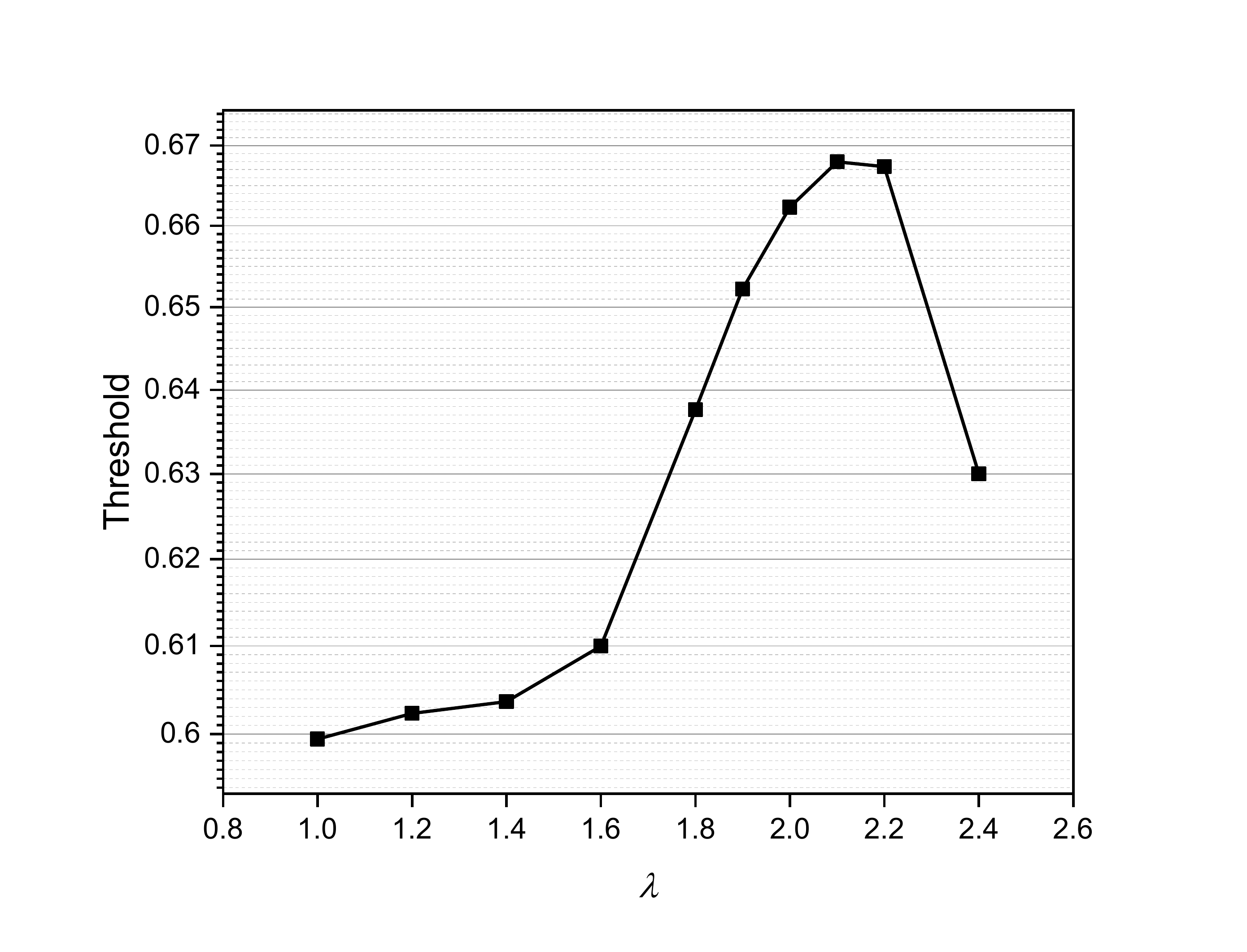}}
\hspace{0.2in}
\subfigure[]{
\label{fig4b}
\includegraphics[width=8.2cm]{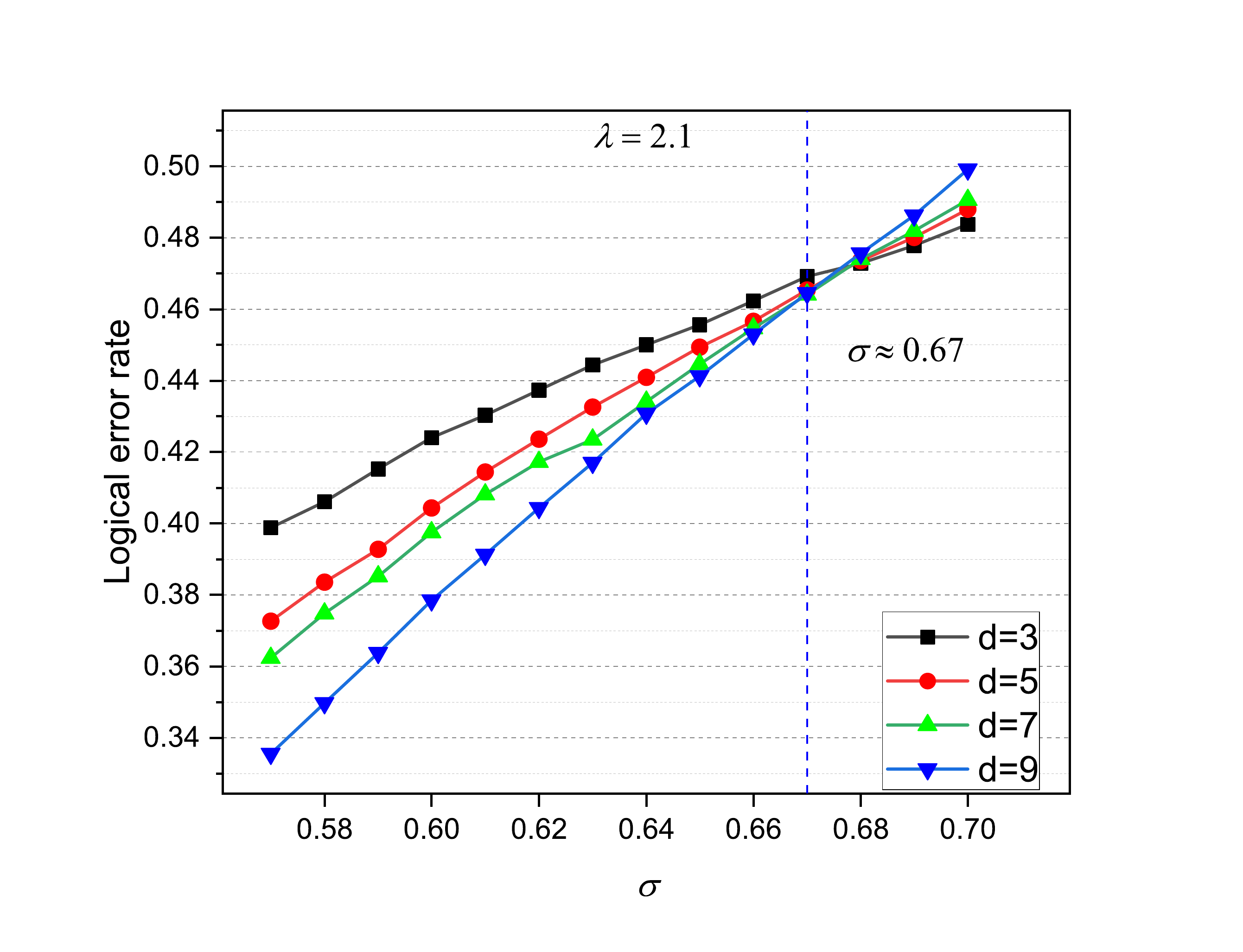}} 
\caption{\justify Code capacity threshold of the XZZX surface-GKP code. (a) Threshold of the  XZZX surface-GKP code as a function of the parameter $\lambda$. With the increasing $\lambda$, the threshold first increases and then decreases, reaching the peak at $\lambda=2.1$. (b) Threshold of the XZZX surface-GKP code at $\lambda=2.1$. The logical error rates with different code distance $d$ are estimated by the Monte Carlo simulation and the maximum threshold of the XZZX surface-GKP code is around 0.67.}
\end{figure*}

\subsection{Code capacity threshold with the designed bias}\label{s3.3}
In this subsection, we show the numerical results of the code capacity threshold of the XZZX surface-GKP code with the designed bias. In the code-capacity noise model, the only noise resource is the Gaussian shift error channel on the data qubits. In other words, all two-qubit gates and measurements are assumed noiseless in both GKP error correction and stabilizer measurement steps. 

We design the bias of the XZZX surface-GKP code by setting different rescaled parameters $\lambda$. Without loss of generality, we set $\lambda\geq1$ such that $\bar{Z}$ error rate is higher than $\bar{X}$ error rate. A $\lambda$ lager than 1 will produce a bias as Eq.$\,$(\ref{e16}) gives. On the other hand, such a $\lambda$ will enlarge the total error rate of $\bar{X}$, $\bar{Y}$ and $\bar{Z}$ error:
\begin{equation}\label{e19}
\bar{p}=1-(1-p_{\bar{X}})(1-p_{\bar{Z}})=p_{\bar{X}}+p_{\bar{Z}}-p_{\bar{X}}p_{\bar{Z}},
\end{equation}
where $p_{\bar{X}}$ and $p_{\bar{Z}}$ are defined in Eq.$\,$(\ref{e15}). The numerical results in Ref.$\,$\cite{bonilla2021xzzx} show that the threshold of the XZZX surface code increases with a larger bias. Therefore, we expect a trade-off that can achieve the optimal threshold of the XZZX surface-GKP code.

To decode the XZZX surface-GKP code, we use the MWPM decoder combined with the continuous-variable information. The mathcing weight in the decoding graph is $w_i=-\log\frac{p_{i}}{1-p_{i}}$, where $p_{i}$ is $p_{\bar{X}}$ or $p_{\bar{Z}}$ of GKP qubit $i$. The numerical results are shown in Fig.$\,$\ref{fig4a} where the optimal threshold is in $\lambda=2.1$ and reaches $\sigma\approx0.67$. Note that when $\lambda=1$, the threshold of XZZX surface code is $\sigma\approx0.60$, almost the same as the standard surface-GKP code \cite{vuillot2019quantum}. When $\lambda=2.1$ and $\sigma$ is in the threshold, one can compute the bias $\eta\approx10^2$ by using Eq.$\,$(\ref{e15}) and Eq.$\,$(\ref{e16}).

A previous work in Ref.$\,$\cite{hanggli2020enhanced} promotes the threshold from $\sigma\approx0.54$ to $\sigma\approx0.58$, where they use the BSV decoder without the GKP continuous-variable information. Compared with that, our threshold is higher and promoted more. We attribute this improvement to the good performance of XZZX surface code under biased noise and the application of the GKP continuous-variable information in the MWPW decoder.

\section{the XZZX surface-GKP code under the full circuit-level noise model}\label{s4}
In this section, we investigate the performance of the XZZX surface-GKP code under the full circuit-level noise model, where the noise in the quantum circuit is taken into account in detail. The section starts with a discussion of the noise model and then proposes the decoding strategy. Lastly, we give the numerical result and compare it with the previous work.

\subsection{Full circuit-level noise model}
The whole quantum circuits for error correction and the time order of the CNOT and CZ gates are illustrated in Fig.$\,$\ref{fig3b}. Here we consider an error model that is as detailed as possible. Specifically, shift errors in $\hat{q}$ and $\hat{p}$ quadratures are assumed to appear after the following operations:

(1) preparations of the initial GKP states (including both data qubits and ancilla qubits),

(2) homodyne measurements of $\hat{q}$ or $\hat{p}$ operator,

(3) idle operations of data qubits when syndrome qubits are measured,

(4) two-qubit gates (including CNOT gates, CZ gates, and balanced beam-splitters).

Simulating the first two kinds of errors is easy. The only thing required is adding a Gaussian shift error channel after each operation in (1-3), as Eq.$\,$(\ref{e14}) given \cite{noh2018quantum}. Let us suppose two Gaussian shift error channels in (1-3) are $\mathcal{N}_p$, $\mathcal{N}_m$ and $\mathcal{N}_i$ with the variance $\sigma_p^2$, $\sigma_m^2$ and $\sigma_i^2$ respectively.

The shift errors after two-qubit gates are more complicated since they produce correlation. In this case, the shift errors on different GKP states cannot be described by independent Gaussian distribution, and we need to introduce the covariance matrices.

Suppose $\mathrm{CNOT}_{\beta}$ gate, $\mathrm{CZ}_{\beta}$ gate and balanced beam-splitter operation are realized by the Hamiltonians $\hat{H}_{CNOT}=g\hat{q_j}\hat{p_k}/\beta$, $\hat{H}_{CZ}=g\hat{q_j}\hat{q_k}/\beta$ and $\hat{H}_{BS}=g\frac{\pi}{4}(\hat{q_j}\hat{p_k}-\hat{p_j}\hat{q_k})$ respectively and GKP states suffer photon loss and heating. Such a noisy gate is equivalent to an ideal gate followed by correlated Gaussian shift errors with the Gaussian distributions $(\hat{q}_j,\hat{q}_k)\sim \boldsymbol N(0,N_{q_jq_k})$ and $(\hat{p}_j,\hat{p}_k)\sim \boldsymbol N(0,N_{p_jp_k})$ (or $(\hat{q}_j,\hat{p}_k)\sim \boldsymbol N(0,N_{q_jp_k})$ and $(\hat{p}_j,\hat{q}_k)\sim \boldsymbol N(0,N_{p_jq_k})$). 
For the $\mathrm{CNOT}_{\beta}$ gate, the covariance matrices are:
\begin{equation}\label{e21}
	\begin{gathered}
	N_{q_jq_k}=\sigma_c^2
	\begin{bmatrix} 1  & \frac{1}{2\beta} \\ \frac{1}{2\beta}  & 1+\frac{1}{3\beta^2} \end{bmatrix},
	\,
	N_{p_jp_k}=\sigma_c^2
	\begin{bmatrix} 1+\frac{1}{3\beta^2} & -\frac{1}{2\beta} \\ -\frac{1}{2\beta} & 1 \end{bmatrix}.
	\end{gathered}
\end{equation}
For the $\mathrm{CZ}_{\beta}$ gate, the covariance matrices are:
\begin{equation}\label{e25}
	\begin{gathered}
	N_{q_jp_k}=\sigma_c^2
	\begin{bmatrix} 1  & \frac{1}{2\beta} \\ \frac{1}{2\beta}  & 1+\frac{1}{3\beta^2} \end{bmatrix},
	\,
	N_{p_jq_k}=\sigma_c^2
	\begin{bmatrix} 1+\frac{1}{3\beta^2} & -\frac{1}{2\beta} \\ -\frac{1}{2\beta} & 1 \end{bmatrix}.
	\end{gathered}
\end{equation}
For the balanced beam-splitter operation, the covariance matrices are:
\begin{equation}\label{e26}
	\begin{gathered}
	N_{q_jq_k}=\sigma_c^2
	\begin{bmatrix} 1 & \quad& 0 \\ 0 & \quad& 1 \end{bmatrix},
	\quad
	N_{p_jp_k}=\sigma_c^2
	\begin{bmatrix} 1 & \quad& 0 \\ 0 & \quad& 1 \end{bmatrix}.
	\end{gathered}
\end{equation}
Here $\sigma_c^2=\kappa/g$, where $g$ is the coupling strength and $\kappa$ is the photon loss and heating rate. The detailed derivations are presented in Appendix \ref{ad}.

\subsection{Maximum likelihood decoding under the full circuit-level noise}\label{4b}
Independently from our work, the ML decoding strategy in GKP error correction is firstly proposed by Ref.$\,$\cite{noh2022low}, in which they only analyze the errors coming from the noisy initial GKP states. Here we generalize their idea to the full circuit-level noise model.

\begin{figure}[t]
\centering
\subfigure[]{
\label{fig5a}
\includegraphics[width=8cm]{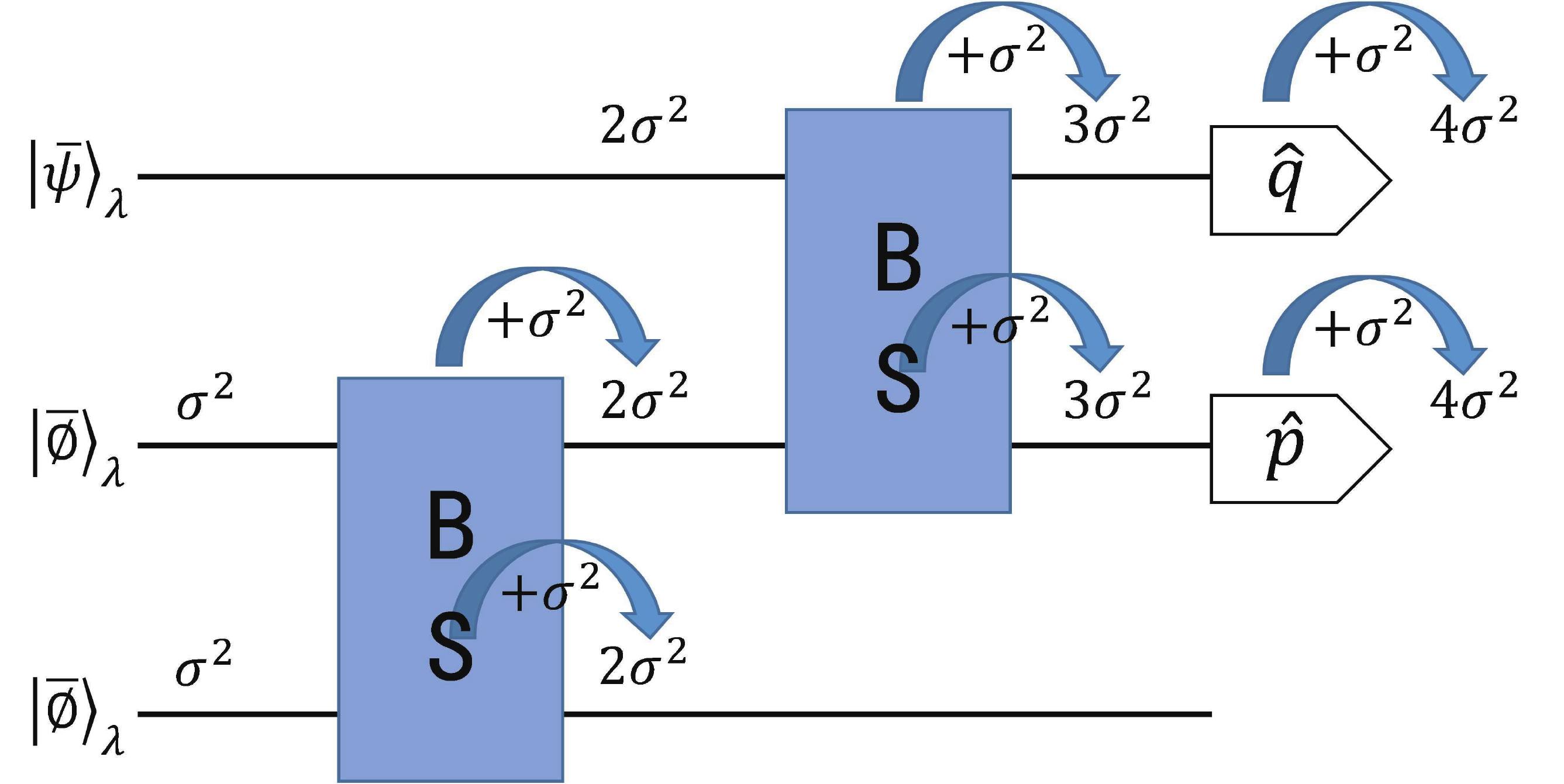}}
\vspace{0.35cm}
\subfigure[]{
\label{fig5b}
\includegraphics[width=9cm]{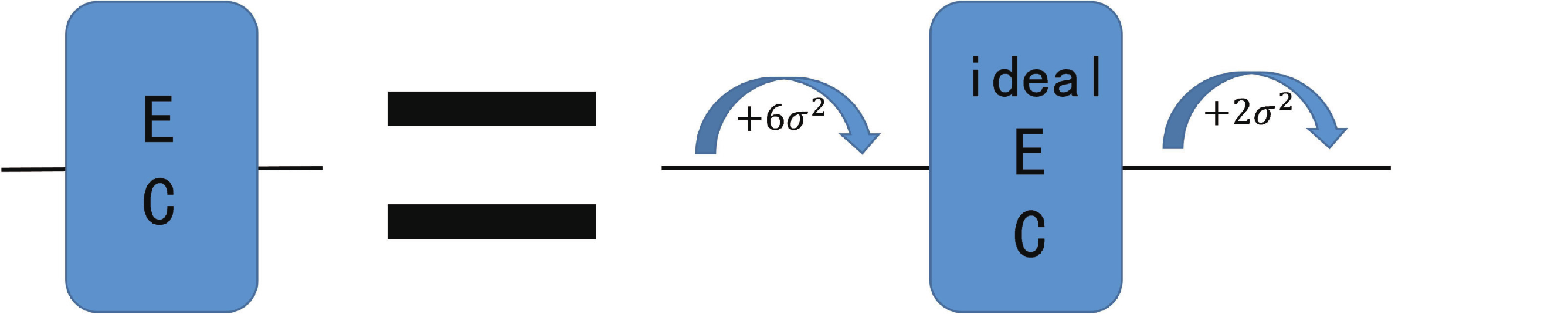}} 
\caption{\justify Shift error propagation analysis in the teleportation-based error correction. (a) Shift errors after each operation. The variances of the shift error after preparations, BS operations, and measurements are assumed equal to $\sigma^2$. The output qubits suffer the shift error with variances $2\sigma^2$. The shift errors in the measurements are $8\sigma^2$ since the Pauli recover operations are determined by $\sqrt{2}q_m$ or $\sqrt{2}p_m$. (b) The circuit identity of the teleportation-based GKP error correction under the full circuit-level noise. The teleportation-based error correction with full circuit-level noise is equivalent to an ideal GKP error correction attached to an input shift error with variance $6\sigma^2$ and an output shift error with variance $2\sigma^2$.}\label{fig5ab}
\end{figure}

To achieve ML decoding after two-qubit gates, it is necessary to analyze the propagation of shift errors in detail. First, let us prove a circuit identity of the teleportation-based GKP error correction under the full circuit-level noise model. In the initial step of the teleportation-based GKP error correction, the data qubit and ancilla qubits suffer the shift errors in the $\hat{q}$ (or $\hat{p}$) quadrature with the variances $\sigma_p^2+\sigma_c^2$ and $\sigma_p^2$ respectively. After the first BS operation, the variances of the errors in ancilla qubits increase to $\sigma_p^2+\sigma_c^2$. Then after the second BS operation, the variances of the errors in the data qubit and the first ancilla qubit increase to $\sigma_p^2+2\sigma_c^2$. Finally, in the measurement step, the variances of the errors in the data qubit and the first ancilla qubit increase to $\sigma_p^2+2\sigma_c^2+\sigma_m^2$ and the results multiply a factor $\sqrt{2}$. Overall, the the variance of the error in the data qubit increase from $\sigma_p^2+\sigma_c^2$ to $2\sigma_p^2+4\sigma_c^2+2\sigma_m^2$, and the output state (the third qubit) suffers the shift error with the variance $\sigma_p^2+\sigma_c^2$. Fig.$\,$\ref{fig5ab} shows an example where we assume that $\sigma\equiv\sigma_p=\sigma_c=\sigma_m=\sigma_i$.

As a result, a noisy GKP error correction by teleportation scheme is equivalent to an ideal GKP error correction attached to an input shift error with variance $\sigma_p^2+3\sigma_c^2+2\sigma_m^2$ and an output shift error with variance $\sigma_p^2+\sigma_c^2$. Furthermore, these two shift errors are independent Gaussian variables because we only apply the orthogonal transformations in three GKP states.

Then let us compute the probability density function of shift errors in $\hat{q}$ quadrature after the $\mathrm{CNOT}_{\beta}$ gate. 
Before the $\mathrm{CNOT}_{\beta}$ gate, the shift errors $\boldsymbol{u}=(u_1,u_2)^T$ of the target and control qubits in $\hat{q}$ quadrature have the covariance matrix:
\begin{equation}\label{e27}
	\begin{gathered}
	N_{q0}=(\sigma_p^2+\sigma_c^2)
	\begin{bmatrix} 1 & 0 \\ 0 & 1 \end{bmatrix},
	\end{gathered}
\end{equation}
which comes from last noisy GKP error correction. After the ideal $\mathrm{CNOT}_{\beta}$ gate, this covariance matrix transforms as follows:
\begin{equation}\label{e28}
	\begin{gathered}
	N_{q}=(\sigma_p^2+\sigma_c^2)
	\begin{bmatrix} 1+\frac{1}{\beta^2} & \frac{1}{\beta} \\ \frac{1}{\beta} & 1 \end{bmatrix}.
	\end{gathered}
\end{equation}
Note that we exchange the order of the control and target qubits since the target qubit is the data qubit and the control qubit is ancilla qubit in the XZZX stabilizer measurement circuits. As mentioned, The shift error comes from the noisy CNOT gate has the covariance matrix:
\begin{equation}\label{e29}
	\begin{gathered}
	N_{c}=\sigma_c^2
	\begin{bmatrix} 1+\frac{1}{3\beta^2}  & \frac{1}{2\beta} \\ \frac{1}{2\beta}  & 1 \end{bmatrix}.
	\end{gathered}
\end{equation}
In the last step, it is required to add the shift errors in the current GKP error correction with the covariance matrix:
\begin{equation}\label{e30}
	\begin{gathered}
	N_{ec}=(\sigma_p^2+3\sigma_c^2+2\sigma_m^2)
	\begin{bmatrix} 1 & 0 \\ 0 & 1 \end{bmatrix}.
	\end{gathered}
\end{equation}
Thus, the total shift errors have the covariance matrix:
\begin{equation}\label{e31}
	\begin{gathered}
	\boldsymbol N= N_q+N_c+N_{ec}.
	\end{gathered}
\end{equation}

\begin{figure}[t]
\centering
\includegraphics[width=8cm]{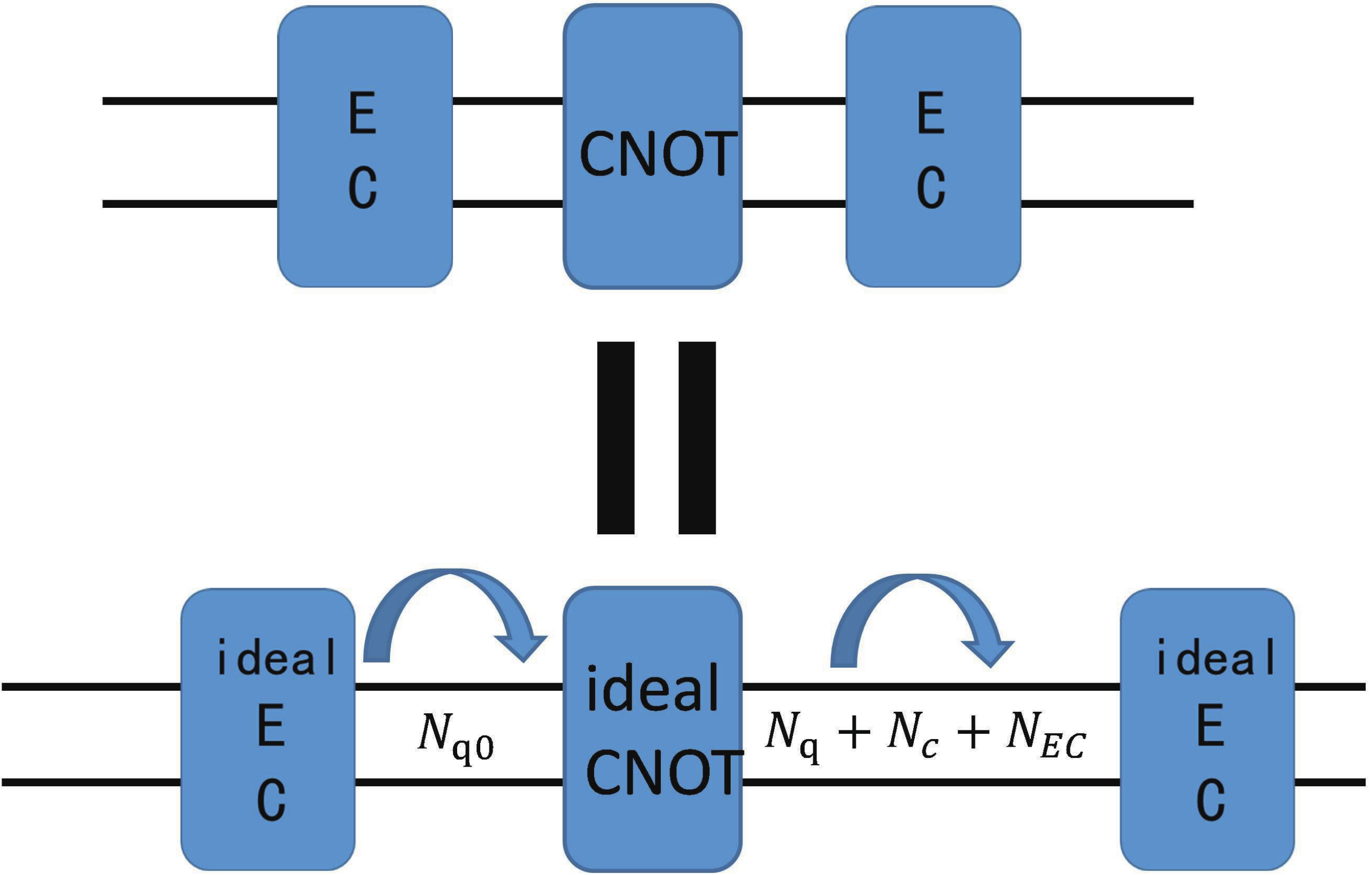}
\caption{\justify The covariances of the shift errors in the noisy circuit. After the first GKP error correction, the output shift errors have the covariance matrix $N_{q0}$. Then after the ideal CNOT gate, the covariance matrix $N_{q0}$ changes to be $N_{q}$. Finally we add the covariance matrices $N_c$ and $N_{EC}$ which come from the noisy CNOT gate and the noisy GKP error correction. Thus, the error before the second GKP error correction decoding has the covariance matrix $N_{q}+N_c+N_{EC}$.}\label{fig add}
\end{figure}

We use Fig.$\,$\ref{fig add} to clearly show the whole process. Accordingly, the probability density function of shift errors in $\hat{q}$ quadrature until an ideal GKP error correction is:
\begin{equation}\label{e32}
\begin{aligned}
&p(u_1,u_2)=\frac{1}{2\pi\sqrt{|\boldsymbol N|}}
\exp[\frac{1}{2}\boldsymbol{u}^T \boldsymbol N ^{-1}\boldsymbol{u}].
\end{aligned}
\end{equation}
Recall that $\boldsymbol{u}=(u_1,u_2)^T$ and $u_1$, $u_2$ are shift errors in $\hat{q}$ quadrature of data qubit and ancilla qubit respectively.

Given the measurement results $q_{m1}$ and $q_{m2}$, the ML decoding process needs to solve the following optimization problem:
\begin{equation}\label{e33}
\begin{aligned}
(n_1,n_2)=\mathop{\arg\min}\limits_{{n_1,n_2}}(\boldsymbol{u}^T \boldsymbol N ^{-1}\boldsymbol{u}),
\end{aligned}
\end{equation}
where $u_1=\sqrt{2}q_{m1}-n_1\sqrt{\pi}$, $u_2=\sqrt{2}q_{m2}-n_2\sqrt{\pi}$. Ref.$\,$\cite{noh2022low} gives the algorithm to solve this kind of problems.

Till now, we have discuss the ML decoding strategy of the shift errors in $\hat{q}$ quadrature after a noisy $\mathrm{CNOT}_{\beta}$ gate. By carrying out a similar derivation, the optimization problem to deal with the shift errors in $\hat{p}$ quadrature after the noisy $\mathrm{CNOT}_{\beta}$ gate is

\begin{equation}\label{e34}
\begin{aligned}
(n'_1,n'_2)=\mathop{\arg\min}\limits_{{n'_1,n'_2}}(\boldsymbol{v}^T \boldsymbol {N} ^{'-1}\boldsymbol{v}),
\end{aligned}
\end{equation}
where $v_1=\sqrt{2}p_{m1}-n'_1\sqrt{\pi}$, $v_2=\sqrt{2}p_{m2}-n'_2\sqrt{\pi}$ and 
\begin{equation}
\begin{gathered}
	\boldsymbol N'= (\sigma_p^2+\sigma_c^2)
	\begin{bmatrix} 1 & -\frac{1}{\beta} \\ -\frac{1}{\beta} & 1+\frac{1}{\beta^2} \end{bmatrix}\\
 +\sigma_c^2
	\begin{bmatrix} 1+\frac{1}{3\beta^2} & \, & -\frac{1}{2\beta} \\ -\frac{1}{2\beta} & \, & 1 \end{bmatrix}+N_{ec}.
	\end{gathered}
\end{equation}
Since the CNOT gate is locally equivalent to the CZ gate, one can get similar results of the $\mathrm{CZ}_{\beta}$ gate by exchanging $\hat{q}$ and $\hat{p}$ operators of the target qubit and repeating the above discussion. As the result, the ML decoding strategy of the $\mathrm{CZ}_{\beta}$ gate needs to optimize the following problem:
\begin{equation}
\begin{aligned}
(n'_1,n_2)=\mathop{\arg\min}\limits_{{n'_1,n_2}}(\boldsymbol{w}^T \boldsymbol N ^{-1}\boldsymbol{w}),\\
(n_1,n'_2)=\mathop{\arg\min}\limits_{{n_1,n'_2}}(\boldsymbol{w'}^{T} \boldsymbol N ^{'-1}\boldsymbol{w'}),
\end{aligned}
\end{equation}
where $\boldsymbol{w}=(v_1,u_2)$, $\boldsymbol{w'}=(u_1$, $v_2)$ and $u_1$, $u_2$, $v_1$, $v_2$ are defined before.

\subsection{Numerical results and overhead estimation}
In this section, we test the logical error rates and thresholds of XZZX surface-GKP codes under three error models by the numerical simulations. In the first error model, we assume that $\sigma\equiv\sigma_p$ and $\sigma_c=\sigma_m=\sigma_i=0$, which corresponds to the situation where the noise is dominated by the squeezing of the GKP states. In the second error model, we assume that $\sigma\equiv\sigma_c=\sigma_m=\sigma_i$ and $\sigma_p=0$, where the noise from two-qubit gates and measurements is dominant. In the third error model, the strengths of these noises are same, i.e., $\sigma\equiv\sigma_c=\sigma_m=\sigma_i=\sigma_p$.

\begin{figure*}[t]
\centering
\subfigure[]{
\label{fig7a}
\includegraphics[width=5.8cm]{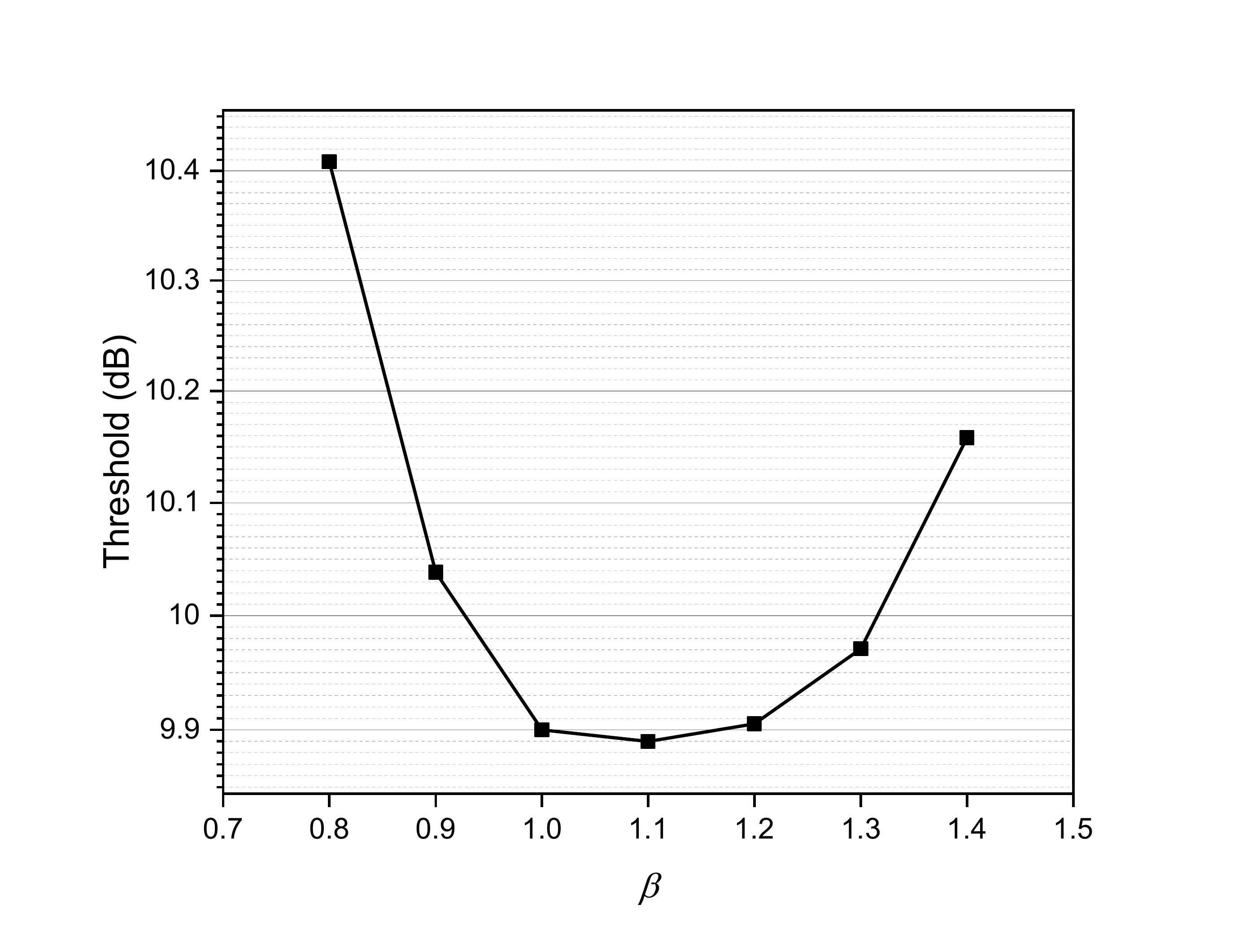}}
\subfigure[]{
\label{fig7b}
\includegraphics[width=5.8cm]{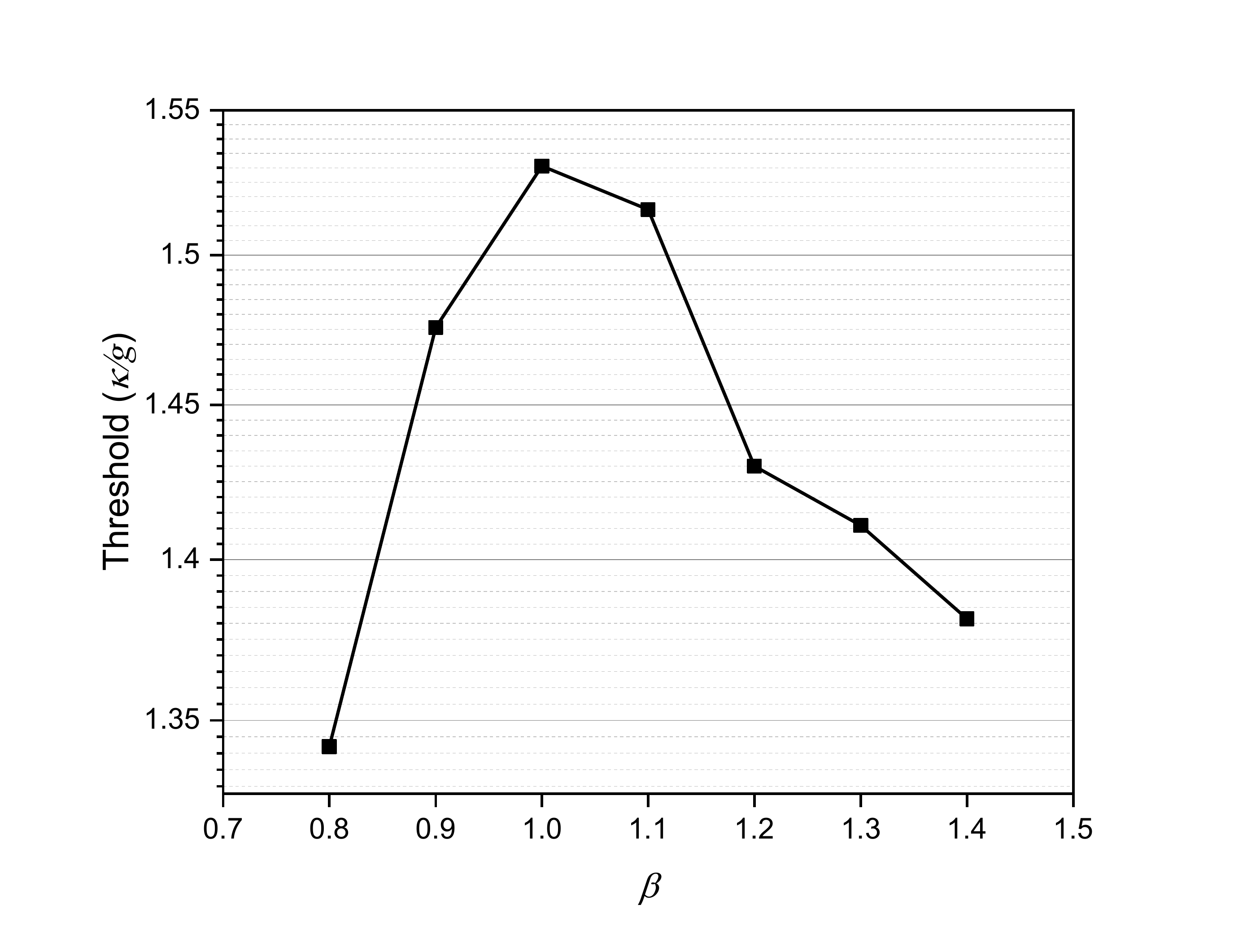}}
\subfigure[]{
\label{fig7c}
\includegraphics[width=5.8cm]{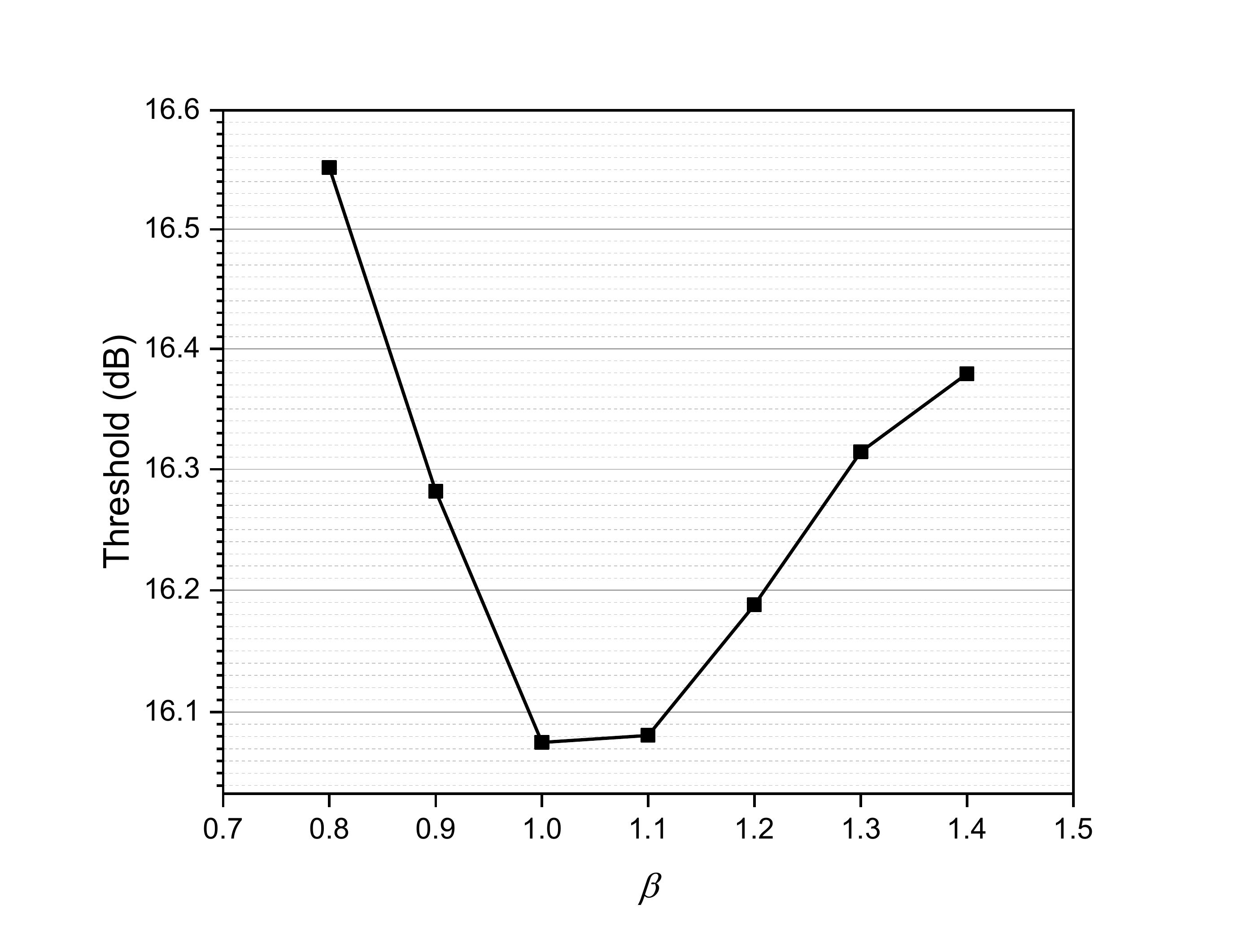}} 
\caption{\justify Thresholds of the XZZX surface-GKP code as a function of the parameter $\beta$ under three circuit-level error models. The optimal thresholds are close to $\beta=1$ in both three cases. The thresholds in the figures are obtained by the intersection points of the logical error rate curves with $d=5$ and $d=7$.}\label{fig7abc}
\end{figure*}

\begin{table*}
\renewcommand{\arraystretch}{2}
\begin{center}   
\caption{The overhead estimation of the XZZX surface-GKP code and the qubit-based surface code.}\label{t1}   
\begin{tabular}{|c|c|}   
\hline   \textbf{XZZX surface-GKP code with $\lambda=1, \beta=1.3$} ($\sigma_{\rm{dB}}=$18.5 dB) & \textbf{qubit-based surface code} ($p=5.64\times 10 ^{-3}$)\\
\hline   $d=3$, $P_L=3.16\times10^{-4}$, 51 GKP states & $d=13$, $P_L=3.63\times10^{-4}$, 337 qubits\\
\hline   $d=5$, $P_L=7.19\times10^{-6}$, 147 GKP states & $d=27$, $P_L=6.59\times10^{-6}$, 1457 qubits\\
\hline   $d=7$, $P_L=2.53\times10^{-7}$, 291 GKP states & $d=39$, $P_L=2.12\times10^{-7}$, 3041 qubits\\ 
\hline   
\end{tabular}   
\end{center}   
\end{table*}

The measurement results of XZZX stabilizers are obtained by homodyne measurements of $\hat{p}$ operator on syndrome qubits, which will be affected by Gaussian shift errors. Thus, in our simulation, we repeat the noisy stabilizer measurement $d$ times appended an ideal stabilizer measurement, where all the components in the circuits are noiseless. Here $d$ is the code distance of the XZZX surface code. As mentioned, we use the ML decoding strategy in the GKP error correction step and use the MWPM decoder to decode XZZX surface code under circuit-level noise model. The GKP continuous-variable information have attached to the matching weights in the decoding graph (see Appendix \ref{ae} for the detail of the matching weights). 

\begin{figure*}[t]
\centering
\subfigure[]{
\label{add2a}
\includegraphics[width=8.2cm]{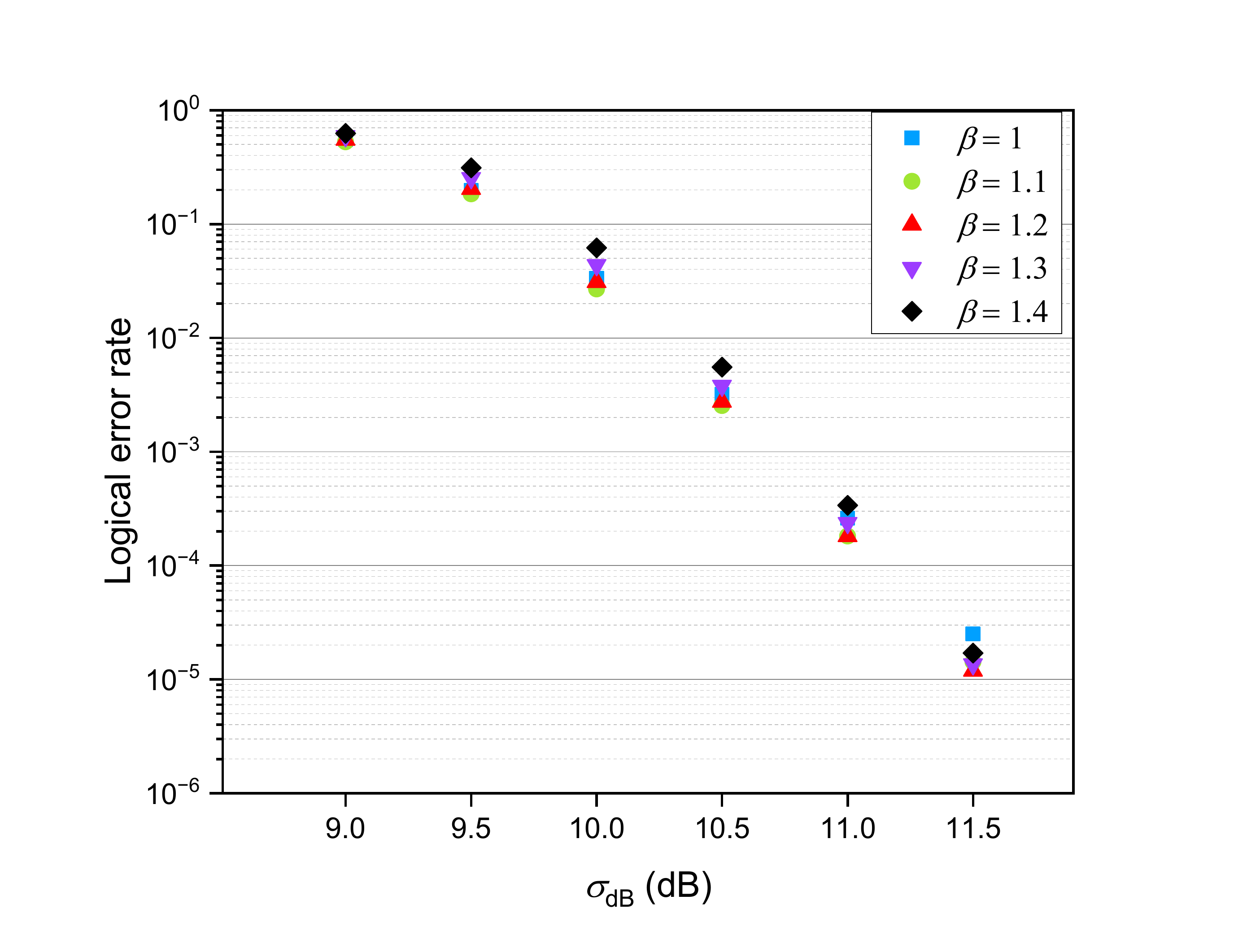}}
\hspace{0.01in}
\subfigure[]{
\label{add2b}
\includegraphics[width=8.2cm]{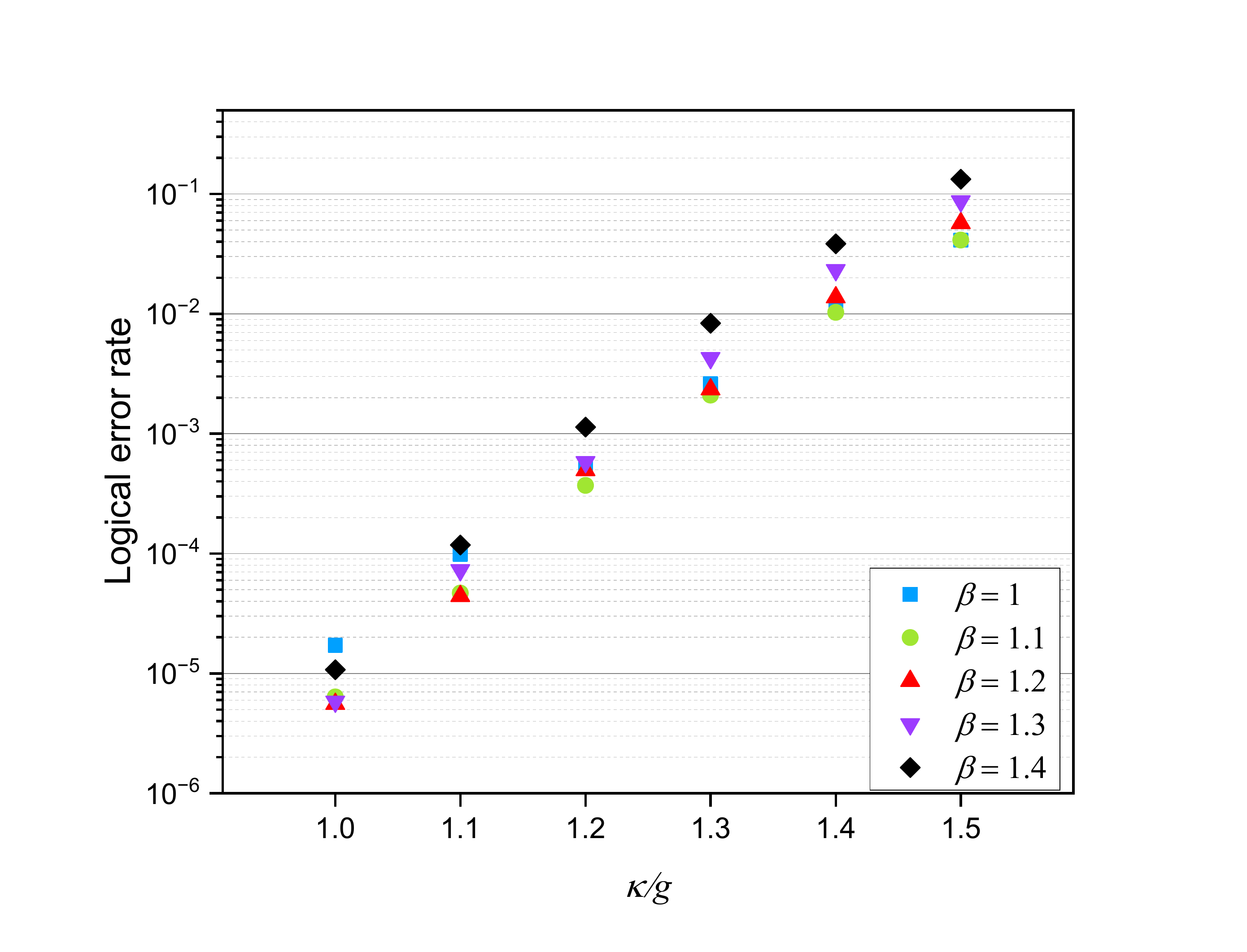}} 
\hspace{0.01in}
\subfigure[]{
\label{add2c}
\includegraphics[width=8.2cm]{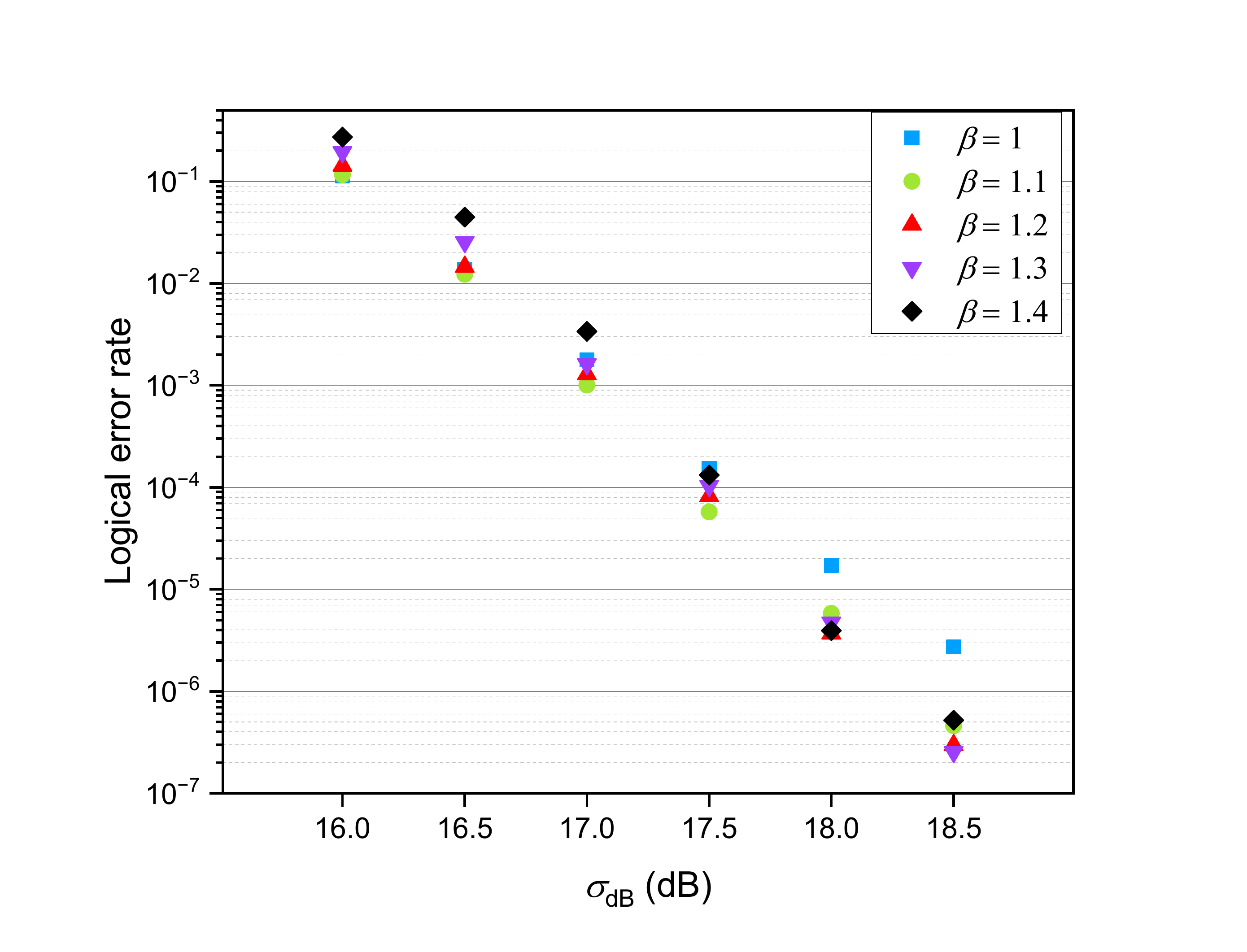}}
\caption{\justify The logical error rates of the XZZX surface-GKP code with $d=7$ and diffrent parameters $\beta$ under three circuit-level error models. When the noise strength decreases, the logical error rates  have been efficiently reduced by asymmetric XZZX surface GKP codes. Within the noise strength the ranges shown in the figures, the logical error rates of the square XZZX surface GKP code and the asymmetric XZZX surface GKP codes differ by at most an order of magnitude. Here we only show the results of $\beta\geq1$ since with the noise strengths shown in the figures, the logical error rates increase when $\beta<1$.}\label{add2}
\end{figure*}

Using the Monte Carlo simulation, we test the thresholds of XZZX surface-GKP code with different parameters $\beta$, where $\lambda$ is fixed to 1. Recall that $\lambda$, $\beta$ are the bias parameters of data qubits and syndrome qubits respectively. We find the optimal thresholds are around $\beta=1$ in both three error models (see Fig.$\,$\ref{fig7abc}). Referring to \cite{noh2022low}, the threshold of the conventional surface-GKP code under the first error model is 9.9 dB, almost equal to that of the XZZX surface-GKP code. Here dB is the unit of the squeezing quantity $\sigma_{\rm{dB}}$ which is defined as $\sigma_{\rm{dB}}=-10\log_{10}(2\sigma^2)$. It seems that the XZZX surface-GKP code provides little improvement. However, when the noise strength is below the threshold, we observe that the logical error rates will be reduced when setting the  appropriate $\beta$. As shown in Fig.$\,$\ref{add2}, the logical error  rates of the XZZX surface-GKP code with distance 7 have been reduced by several times when setting $\beta>1$ in some cases. For example, under the third error model with $\sigma_{\rm{dB}}=$ 18.5dB ($\kappa/g = 0.706\%$), the logical error rate of the XZZX surface-GKP code with $\lambda=1$, $\beta=1.3$ has decreased by an order of magnitude  compared with the square XZZX surface-GKP code. In the previous work \cite{noh2022low}, this phenomenon only occurs when $\lambda=1$, $\beta=1.1$ in the first error model and the logical error rates are only redeced by only a small fraction. These results indicate the XZZX surface-GKP codes are more suitable for asymmetric concatenation under the general error models.

Meanwhile, as a complement to the previous work, Fig.$\,$\ref{fig7abc} also shows the threshold of square XZZX surface-GKP code ($\beta=1$) under the second and third error model using the ML decoding. The thresholds are about $\kappa/g \approx 1.53\%$ and $\sigma_{\rm{dB}}\approx$ 16.1 dB ($\kappa/g \approx 1.23\%$) for the second and third error models respectively. These results of the conventional surface-GKP code  are $\kappa/g \approx 0.81\%$ and $\sigma_{\rm{dB}}\approx$ 18.6 dB ($\kappa/g \approx 0.69\%$) without using the ML decoding strategy \cite{noh2020fault}. The improvements demonstrate the effectiveness of the ML decoding under the full circuit-level noise models.

Furthermore, we estimate the overhead of the XZZX surface-GKP code to reach a low logical error rate $P_L$ under the third error model and compare it with the overhead of the qubit-based surface code. In order to encode a logical qubit with code distance $d$, an XZZX surface-GKP code needs totally $3(2d^2-1)$ GKP states ($d^2$ data qubits with $2d^2$ GKP ancilla qubits and $d^2-1$ syndrome qubits with $2(d^2-1)$ GKP ancilla qubits), while a qubit-based surface code requires $2d^2-1$ qubits. Through the optimization of the quantum circuits in the future, some GKP qubits may be reused to reduce the total qubit consumption further. Here we only provide a pessimistic estimation of the overhead of the XZZX surface-GKP code. 

The overhead of the qubit-based surface code is estimated by the empirical formula $P_L\cong 0.01(100p)^{\frac{d+1}{2}}$. Note that the empirical formula is obtained from a circuit-level depolarizing noise model \cite{fowler2012surface}, and $P_L$ is only the logical error rate of $X_L$ or $Z_L$ which only corresponds to half of $P_L$ considered in the XZZX surface-GKP code. In order to fit the empirical formula, we test the error probability of two qubit-gates, preparations, measurements and idles in the XZZX surface-GKP code. The result shows that when $\sigma_{\rm{dB}}=$ 18.5 dB ($\kappa/g \approx 0.71\%$), the errors occur with probability $p\approx 5.64\times10^{-3}$ after two qubit-gates which is much higher than other operations with $p'\approx 1.9\times10^{-4}$. Hence, we only consider the main term $p\approx 5.64\times10^{-3}$ as the qubit error rate in the empirical formula.

The whole comparison is shown in Table.$\,$\ref{t1}. For example, the overhead to reach $P_L=2.53\times10^{-7}$ by the XZZX surface-GKP code requires the code with the $d=7$ and $\lambda=1$, $\beta=1.3$. In contrast, the qubit-encoded surface code to reach a close $P_L$ needs $d=39$, which is much larger than XZZX surface-GKP code.

\section{Conclusion and outlook}\label{s5}
In this paper, we study the concatenation of the GKP code with the XZZX surface code. By designing bias from the rectangular GKP code, the threshold of the XZZX surface-GKP code is improved under the code-capacity noise model. In addition, our paper also considers more realistic noise models and provides the ML decoding strategy. The numerical results show that, in some cases, the threshold and the logical error rates outperform those in the previous work. Lastly, we analyze the advantages of the XZZX surface-GKP code from the perspective of overhead compared with the qubit-based surface code.

In the code-capacity noise model, we introduce the bias by designing the asymmetric GKP code and witness a significant increase in the threshold. However, in the circuit-level noise model, the threshold is barely improved by the asymmetric GKP codes. An interesting open question is whether exists other effective way in the circuit-level noise model to introduce bias. Meanwhile, we expect the realization of bias-preserving gates of the GKP code, just like the previous work in the cat code \cite{puri2020bias,guillaud2019repetition}. By that time, using the array of a rectangle XZZX surface code with $d_z\approx d_x(1-\frac{\log \eta}{\log p})$ \cite{bonilla2021xzzx} will further reduce the overhead.

In summary, we believe that the XZZX GKP-surface code is a competitive candidate for large-scaled fault-tolerant quantum computation, and is excepted a brilliant future for the experimental realization.  

\acknowledgments
This work was supported by the National Natural Science Foundation of China (Grant No. 12034018) and Innovation Program for Quantum Science and Technology (Grant No. 2021ZD0302300).

\appendix
\section{The GKP error correction and the XZZX stabilizer measurement}
\subsection{The detail of the teleportation-based error correction scheme}\label{aa}
Let us show how the teleportation-based error correction scheme works. Firstly, a Bell state is produced by two qunaught states with the action of the balanced beam-splitter:
\begin{equation}
\begin{aligned}
&B_{\frac{\pi}{4}}\ket{\emptyset_\lambda}\ket{\emptyset_\lambda}\\
\propto& B_{\frac{\pi}{4}}\sum_{m\in\mathbb{Z}}\ket{\hat{q}=m\sqrt{2\pi}\lambda}
\sum_{n\in\mathbb{Z}}\ket{\hat{q}=n\sqrt{2\pi}/\lambda}\\
\propto& \sum_{m,n}\ket{\hat{q}=(m-n)\sqrt{\pi}\lambda}
\ket{\hat{q}=(m+n)\sqrt{\pi}/\lambda}\\
\propto& \sum_{m,n\, is\, even}\ket{\hat{q}=m'\sqrt{\pi}\lambda}
\ket{\hat{q}=n'\sqrt{\pi}/\lambda}+\\
&\sum_{m,n\, is\, odd}\ket{\hat{q}=m'\sqrt{\pi}\lambda}
\ket{\hat{q}=n'\sqrt{\pi}/\lambda}\\
\propto& \quad |\bar{0}_\lambda\rangle|\bar{0}_\lambda\rangle+|\bar{1}_\lambda\rangle|\bar{1}_\lambda\rangle.
\end{aligned}
\end{equation}
Since $B_{\frac{\pi}{4}}$ also is a real unitary operator, that is, $B_{\frac{\pi}{4}}=B^{-1}_{\frac{\pi}{4}}$, we have:
\begin{equation}
\begin{aligned}
 B_{\frac{\pi}{4}} (|\bar{0}_\lambda\rangle|\bar{0}_\lambda\rangle+|\bar{1}_\lambda\rangle|\bar{1}_\lambda\rangle) \propto \ket{\emptyset_\lambda}\ket{\emptyset_\lambda}.
\end{aligned}
\end{equation}
Using Eq.$\,$(\ref{e13}), one can obtain:
\begin{equation}
\begin{aligned}
 &B_{\frac{\pi}{4}} (|\bar{0}_\lambda\rangle|\bar{1}_\lambda\rangle+|\bar{1}_\lambda\rangle|\bar{0}_\lambda\rangle) \\
 =&B_{\frac{\pi}{4}}e^{-i\hat{p}_1\sqrt{\pi}\lambda} (|\bar{0}_\lambda\rangle|\bar{0}_\lambda\rangle+|\bar{1}_\lambda\rangle|\bar{1}_\lambda\rangle)\\
 \propto&  e^{-i\hat{p}_1\sqrt{\pi/2}\lambda}e^{-i\hat{p}_2\sqrt{\pi/2}\lambda}\ket{\emptyset_\lambda}\ket{\emptyset_\lambda},\\
 &B_{\frac{\pi}{4}} (|\bar{0}_\lambda\rangle|\bar{0}_\lambda\rangle-|\bar{1}_\lambda\rangle|\bar{1}_\lambda\rangle) \\
 \propto&  e^{-i\hat{q}_1\sqrt{\pi/2}\lambda}e^{-i\hat{q}_2\sqrt{\pi/2}\lambda}\ket{\emptyset_\lambda}\ket{\emptyset_\lambda},\\
  &B_{\frac{\pi}{4}} (|\bar{0}_\lambda\rangle|\bar{1}_\lambda\rangle-|\bar{1}_\lambda\rangle|\bar{0}_\lambda\rangle) \\
 \propto&  e^{-i\hat{p}_1\sqrt{\pi/2}\lambda}e^{-i\hat{p}_2\sqrt{\pi/2}\lambda}e^{-i\hat{q}_1\sqrt{\pi/2}\lambda}e^{-i\hat{q}_2\sqrt{\pi/2}\lambda}\ket{\emptyset_\lambda}\ket{\emptyset_\lambda}.
\end{aligned}
\end{equation}
Then, we can achieve the Bell basis measurement by measuring $\hat{q}_1$ and $\hat{p}_2$.

The input state  has the following decomposition:
\begin{equation}
\begin{aligned}
&\ket{\bar{\psi}_\lambda}
(\ket{\bar{0}_\lambda}
\ket{\bar{0}_\lambda}+
\ket{\bar{1}_\lambda}
\ket{\bar{1}_\lambda})\\=
&(a\ket{\bar{0}_\lambda}+
b\ket{\bar{1}_\lambda})
(\ket{\bar{0}_\lambda}
\ket{\bar{0}_\lambda}+
\ket{\bar{1}_\lambda}
\ket{\bar{1}_\lambda})\\=
&(\ket{\bar{0}_\lambda}
\ket{\bar{0}_\lambda}+
\ket{\bar{1}_\lambda}
\ket{\bar{1}_\lambda})
(a\ket{\bar{0}_\lambda}+
b\ket{\bar{1}_\lambda})\\+
&(\ket{\bar{0}_\lambda}
\ket{\bar{0}_\lambda}-
\ket{\bar{1}_\lambda}
\ket{\bar{1}_\lambda})
(a\ket{\bar{0}_\lambda}-
b\ket{\bar{0}_\lambda})\\+
&(\ket{\bar{0}_\lambda}
\ket{\bar{1}_\lambda}+
\ket{\bar{1}_\lambda}
\ket{\bar{0}_\lambda})
(a\ket{\bar{1}_\lambda}+
b\ket{\bar{0}_\lambda})\\+
&(\ket{\bar{0}_\lambda}
\ket{\bar{1}_\lambda}-
\ket{\bar{1}_\lambda}
\ket{\bar{0}_\lambda})
(a\ket{\bar{1}_\lambda}-
b\ket{\bar{0}_\lambda}),
\end{aligned}
\end{equation}
where $\ket{\bar{\psi}_\lambda}=a\ket{\bar{0}_\lambda}+
b\ket{\bar{1}_\lambda}$. After the balanced beam-splitter on the first two qubits, the four Bell states will be distinguished by the measurement results $q_m$ and $p_m$. The final state of the third qubit is $\bar{X}^{n_1}\bar{Z}^{n'_1}\ket{\bar{\psi}_\lambda}$, where $n_1=\sqrt{2}q_m\lambda/\sqrt{\pi}$ and $n'_1=\sqrt{2}p_m/\sqrt{\pi}/\lambda$. If we consider the Gaussian shift errors on the input states, $n_1$ and $n'_1$ typically are not integers. The simple closest-integer decoding gives $n_1=\lfloor \frac{\sqrt{2}q_m\lambda}{\sqrt{\pi}}+\frac{1}{2}\rfloor$ and $n'_1=\lfloor \frac{\sqrt{2}p_m}{\sqrt{\pi}\lambda}+\frac{1}{2}\rfloor$. However, after acting the two-qubit gate, such as CNOT or CZ gate, the closest-integer decoding is no longer optimal. The optimal $n_1$ and $n'_1$ are obtained by the ML decoding as presented in the main text.

\subsection{The superiority of the teleportation based scheme over the Steane scheme}\label{ab}
\begin{figure}[b]
\centering
\includegraphics[width=9cm]{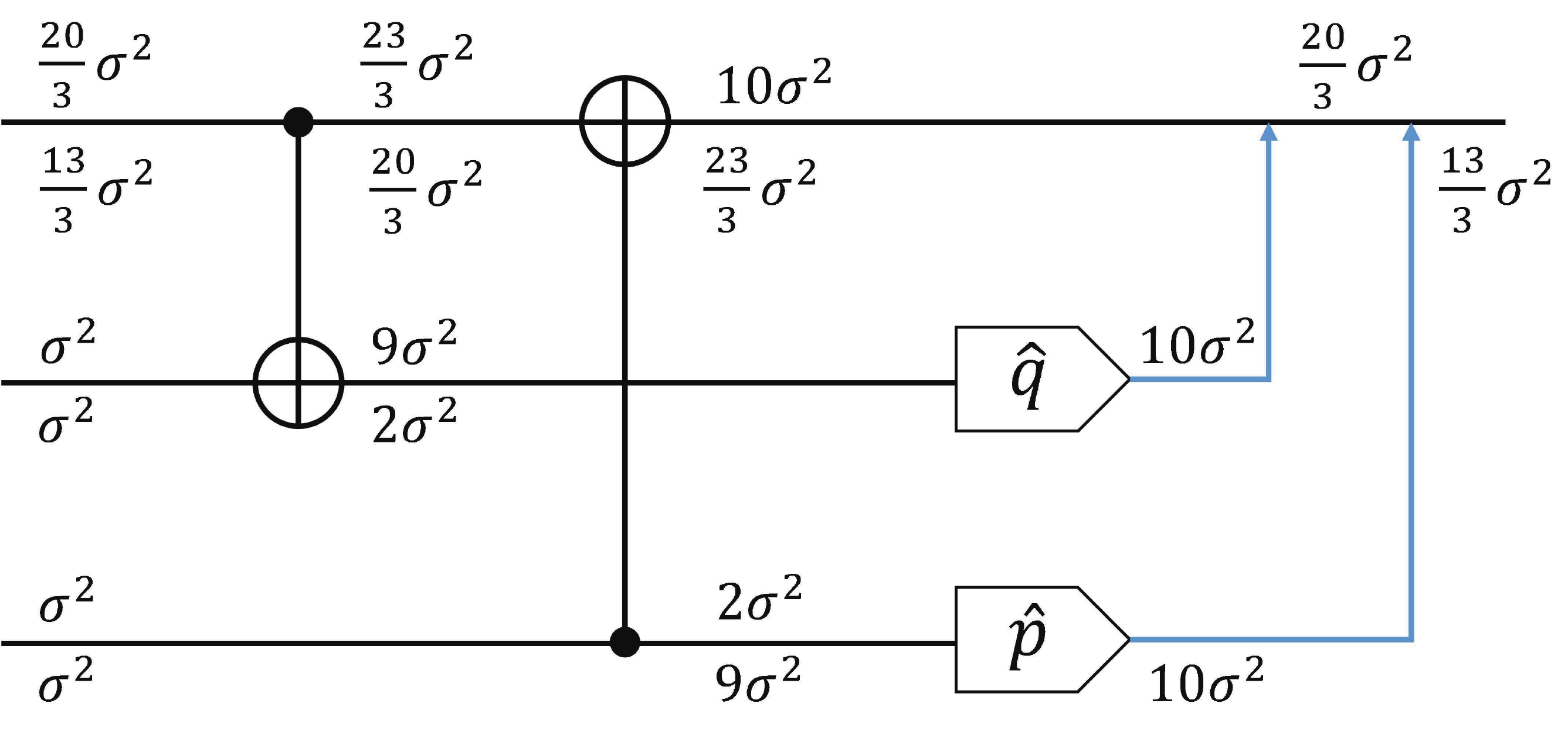}
\caption{\justify Shift error propagation analysis of the Steane error correction under full circuit-level noise. The final data qubits suffers the shift errors with variances $\frac{20}{3}\sigma^2$ in $\hat{q}$ quadrature (above the lines) and $\frac{13}{3}\sigma^2$ in $\hat{p}$ quadrature (below the lines).  The shift errors in the measurements are $10\sigma^2$ in both $\hat{q}$ and $\hat{p}$ quadratures.}\label{fig7}
\end{figure}

In Fig.$\,$\ref{fig7}, we analyse the error propagation in the Steane scheme where we also assume that $\sigma\equiv\sigma_p=\sigma_c=\sigma_m$. In the $\hat{q}$ quadrature, the initial data qubit suffers the error $u$ with the variances $\frac{20}{3}\sigma^2$ which comes from the last GKP error correction (see below for explanation).  Let $u_1,u_2$ be initial errors in the ancila qubits and $u_{c1},u'_{c1},u_{c2},u'_{c2}$ be the errors after each CNOT gate. After the first CNOT gate, the errors in the data qubit and the first ancilla qubit are $u+u_{c1}$ with the variance $\frac{20}{3}\sigma^2+\sigma^2=\frac{23}{3}\sigma^2$ and $u_1+u+u'_{c1}$ with the variance $\sigma^2+\frac{20}{3}\sigma^2+\frac{4}{3}\sigma^2=9\sigma^2$ respectively. After the second CNOT gate, the error of data qubit increases to $u+u_{c1}+u_2+u_{c2}$ with the variance $10\sigma^2$. In addition to the error $u_m$ with variance $\sigma^2$ in the measurement, the final error of the measurement result is $u_1+u+u'_{c1}+u_m$ with the variance $10\sigma^2$. Thus, the error in the output qubits are $(u+u_{c1}+u_2+u_{c2})-(u_1+u+u'_{c1}+u_m)$ with the variance $\frac{20}{3}\sigma^2$.

The discussion in the $\hat{p}$ quadrature is similar. The result shows that the errors after the error correction have the variances $\frac{20}{3}\sigma^2$ and $\frac{13}{3}\sigma^2$ in $\hat{q}$ and $\hat{p}$ quadratures respectively, which is higher than those in the teleportation based scheme. On the other hand, the error before the measurement has the variances $10\sigma^2$, leading to a higher logical error probability than the teleportation based scheme. Thus, from these two perspectives, the teleportation-based error correction scheme is superior to the Steane scheme.

\subsection{The XZZX stabilizer measurement circuit}\label{ac}
The stabilizer measurement is projecting the state $\ket{\psi}$ to the eigenspace of the XZZX stabilizer. The key step is the action of four CNOT or CZ gates:
\begin{equation}
\begin{aligned}
\ket{0}\ket{\psi}+\ket{1}XZZX\ket{\psi},
\end{aligned}
\end{equation}
where the first qubit is the syndrome qubit.
If $XZZX$ stabilizes $\ket{\psi}$, the final state is $\ket{+}\ket{\psi}$. If $XZZX\ket{\psi}=-\ket{\psi}$, the final state is $\ket{-}\ket{\psi}$. Hence, the measurement on the first qubit will give the syndrome information.

Note that now our qubits are rectangle GKP states. Specifically, the data qubit is $\ket{\bar{\psi}_\lambda}$ and the syndrome qubit is $\ket{\bar{+}_\beta}$. In the stabilizer measurement circuit, the rescaled CNOT gate $\mathrm{CNOT}_{\beta/\lambda}=e^{-i\hat{q}_1\hat{p}_2\lambda/\beta}$ performs $\bar{X}_\lambda=e^{-i(\sqrt{\pi}\lambda)\hat{p}_2}$ on the data qubit if ${q}_1=(2k+1)\sqrt{\pi}\beta$. Likewise, the rescaled CZ gate $\mathrm{CZ}_{\lambda\beta}=e^{-i\hat{q}_1\hat{q}_2/(\lambda\beta)}$ performs $\bar{Z}_\lambda=e^{-i(\sqrt{\pi}/\lambda)\hat{q}_2}$ on the data qubit if ${q}_1=(2k+1)\sqrt{\pi}\beta$.

The parameters $\lambda$ and $\beta$ are directly affects the error rates and the error propagation. For instance, a large $\beta$ will lead to a high $\bar{Z}_\beta$ error rate on the syndrome qubit. However, a large $\beta$ will also reduce the propagated error since $\mathrm{CNOT}_{\beta/\lambda}e^{-iup_1}
=e^{-iu\hat{p}_1}e^{-iu\hat{p}_2\lambda/\beta}
\mathrm{CNOT}_{\beta/\lambda}$  and
$\mathrm{CZ}_{\lambda\beta}e^{-iu\hat{p}_1}
=e^{-iu\hat{p}_1}e^{iu\hat{q}_2/(\lambda\beta)}
\mathrm{CZ}_{\lambda\beta}$.

\section{Derivations of the covariance matrices for the noisy gates}\label{ad}

Although beam-spliter operations can be realized directly in optical systems, it is more general to implement beam-spliter operations by engineering time-dependent couplings \cite{gao2018programmable,zhang2019engineering}. Therefore, in the main text, we assume $\mathrm{CNOT}_{\beta}$ gate, $\mathrm{CZ}_{\beta}$ gate and balanced beam-splitter operation are realized by the Hamiltonians $\hat{H}_{CNOT}=g\hat{q_j}\hat{p_k}/\beta$, $\hat{H}_{CZ}=g\hat{q_j}\hat{q_k}/\beta$ and $\hat{H}_{BS}=g\frac{\pi}{4}(\hat{q_j}\hat{p_k}-\hat{p_j}\hat{q_k})$ respectively and GKP states suffer photon loss and heating.  This evolution can be described by $\exp[\mathcal{L}t]$ where $\mathcal{L}$ has the form:
\begin{equation}\label{d1}
\begin{aligned}
\mathcal{L}(\hat{\rho})
=&-i[\hat{H},\hat{\rho}]
+\mathcal{D}(\hat{\rho})\\
=&-i[\hat{H},\hat{\rho}]
+\kappa(\hat{a}_j\hat{\rho}\hat{a}_j^\dag-\frac{1}{2}\hat{a}_j^\dag\hat{a}_j\hat{\rho}-\frac{1}{2}\hat{\rho}\hat{a}_j^\dag\hat{a}_j)\\
&+\kappa(\hat{a}_k\hat{\rho}\hat{a}_k^\dag-\frac{1}{2}\hat{a}_k^\dag\hat{a}_k\hat{\rho}-\frac{1}{2}\hat{\rho}\hat{a}_k^\dag\hat{a}_k).
\end{aligned}
\end{equation}
Here $\kappa$ is the photon loss and heating rate.

Utilizing Trotter decomposition $\exp(-i\hat{V}t)= \lim_{n\rightarrow\infty}(\exp(-i\hat{V}t/n))^n$          \cite{trotter1959product}, the gate operations are realized by repeating $\exp(-i\hat{H}\Delta t/N)$ operations $N$ times, where $N$ is large enough, $\Delta t=1/g$ and $\hat{H}$ is $\hat{H}_{CNOT}$, $\hat{H}_{CZ}$, or $\hat{H}_{BS}$. Note that by Eq.$\,$(\ref{e10}) and Eq.$\,$(\ref{e13}), $\exp(-\hat{q_j}\hat{p_k}/\beta /N)$, $\exp(-\hat{q_j}\hat{q_k}/\beta /N)$ and $\exp[\frac{\pi}{4}(\hat{q_j}\hat{p_k}-\hat{p_j}\hat{q_k})/N]$ transforms  $(\hat{q_j},\hat{q_k}, \hat{p_j},\hat{p_k})^T$ by the following matrices:
\begin{equation}
\boldsymbol T_{q_jq_k}^{^{CNOT}}=
\begin{bmatrix} 1 & \quad &0 \\ 
\frac{1}{\beta N} &\quad &1 \end{bmatrix},  
\boldsymbol T_{p_jp_k}^{^{CNOT}}=
\begin{bmatrix} 1 & \quad &-\frac{1}{\beta N} \\ 
0 &\quad &1 \end{bmatrix}; 
\end{equation}\label{d2}

\begin{equation}
\boldsymbol T_{q_jp_k}^{^{CZ}}=
\begin{bmatrix} 1 & \quad &0 \\ 
\frac{1}{\beta N} &\quad &1 \end{bmatrix},  
\boldsymbol T_{p_jq_k}^{^{CZ}}=
\begin{bmatrix} 1 & \quad &-\frac{1}{\beta N} \\ 
0 &\quad &1 \end{bmatrix}; 
\end{equation}\label{d3}

\begin{equation}
\begin{aligned}
\boldsymbol T_{q_jq_k}^{^{BS}}=
\begin{bmatrix} \cos(\frac{\pi}{4N})  &-\sin(\frac{\pi}{4N}) \\ 
\sin(\frac{\pi}{4N})  &\cos(\frac{\pi}{4N}) \end{bmatrix}, \\
\,\\
\boldsymbol T_{p_jp_k}^{^{BS}}=
\begin{bmatrix} \cos(\frac{\pi}{4N})  &-\sin(\frac{\pi}{4N}) \\ 
\sin(\frac{\pi}{4N})  &\cos(\frac{\pi}{4N}) \end{bmatrix}.
\end{aligned}
\end{equation}\label{d4}

After realizing $\exp(-i\hat{H}\Delta t/N)$ in one time step (say $n$th step), the photon loss error term $\mathcal{D}(\hat{\rho})$ in Eq.$\,$(\ref{d1}) leads to Gaussian shift errors with the covariance matrix\cite{noh2020fault,noh2018quantum}:
\begin{equation}\label{e23}
	\begin{gathered}
	N_{q,n}=\frac{\sigma_c^2}{N}
	\begin{bmatrix} 1 & \quad& 0 \\ 0 & \quad& 1 \end{bmatrix},
	\end{gathered}
	\begin{gathered}
	N_{p,n}=\frac{\sigma_c^2}{N}
	\begin{bmatrix} 1 & \quad& 0 \\ 0 & \quad& 1 \end{bmatrix},
	\end{gathered}
\end{equation}
where $\sigma_c^2=\kappa\Delta t=\kappa/g$.
After the later $N-n$ steps, these covariance matrices will transform to $(\boldsymbol T_q)^{N-n}N_{q,n}(\boldsymbol T_q^T)^{N-n}$ and $(\boldsymbol T_p)^{N-n}N_{p,n}(\boldsymbol T_p^T)^{N-n}$. where for the CNOT gate, $\boldsymbol T_q=\boldsymbol T_{q_jq_k}^{^{CNOT}}$  and $\boldsymbol T_p=\boldsymbol T_{p_jp_k}^{^{CNOT}}$      As the result, for the CNOT gate, the total shift errors after $N$ time steps have the covariance matrix:
\begin{equation}
	\begin{gathered}
	N_{q_jq_k}^{^{CNOT}}=\lim_{N\rightarrow\infty}\sum_{n=1}^{N}(\boldsymbol T_q)^{N-n}N_{q,n}(\boldsymbol T_q^T)^{N-n}
	\\=\sigma_c^2
	\begin{bmatrix} 1 & \quad & \frac{1}{2\beta} \\ \frac{1}{2\beta} & \quad & 1+\frac{1}{3\beta^2} \end{bmatrix},
	\end{gathered}
\end{equation}

\begin{equation}
	\begin{gathered}
	N_{p_jp_k}^{^{CNOT}}=\lim_{N\rightarrow\infty}\sum_{n=1}^{N}(\boldsymbol T_p)^{N-n}N_{q,n}(\boldsymbol T_p^T)^{N-n}
	\\=\sigma_c^2
	\begin{bmatrix} 1+\frac{1}{3\beta^2} & -\frac{1}{2\beta} \\ -\frac{1}{2\beta} & 1 \end{bmatrix}.
	\end{gathered}
\end{equation}

Likewise, one can easily derive the covariance matrices of the shift error after the CZ gate and balanced beam-splitter by using their own transformation matrices $\boldsymbol T_{q_jp_k}^{^{CZ}},\boldsymbol T_{p_jq_k}^{^{CZ}},\boldsymbol T_{q_jq_k}^{^{BS}},\boldsymbol T_{p_jp_k}^{^{BS}}$. The results are stated as follows:
\begin{equation}\label{d25}
	\begin{gathered}
	N_{q_jp_k}^{^{CZ}}=\sigma_c^2
	\begin{bmatrix} 1 & \quad & \frac{1}{2} \\ \frac{1}{2} & \quad & 1+\frac{1}{3\beta^2} \end{bmatrix},
	\,
	N_{p_jq_k}^{^{CZ}}=\sigma_c^2
	\begin{bmatrix} 1+\frac{1}{3\beta^2} & -\frac{1}{2\beta} \\ -\frac{1}{2\beta} & 1 \end{bmatrix};
	\end{gathered}
\end{equation}
\begin{equation}\label{d26}
	\begin{gathered}
	N_{q_jq_k}^{^{BS}}=\sigma_c^2
	\begin{bmatrix} 1 & \quad& 0 \\ 0 & \quad& 1 \end{bmatrix},
	\quad
	N_{p_jp_k}^{^{BS}}=\sigma_c^2
	\begin{bmatrix} 1 & \quad& 0 \\ 0 & \quad& 1 \end{bmatrix}.
	\end{gathered}
\end{equation}

\section{Matching weights  of the MWPM decoder}\label{ae}
\begin{figure*}
\centering
\subfigure[]{
\label{fig90}
\includegraphics[width=5cm]{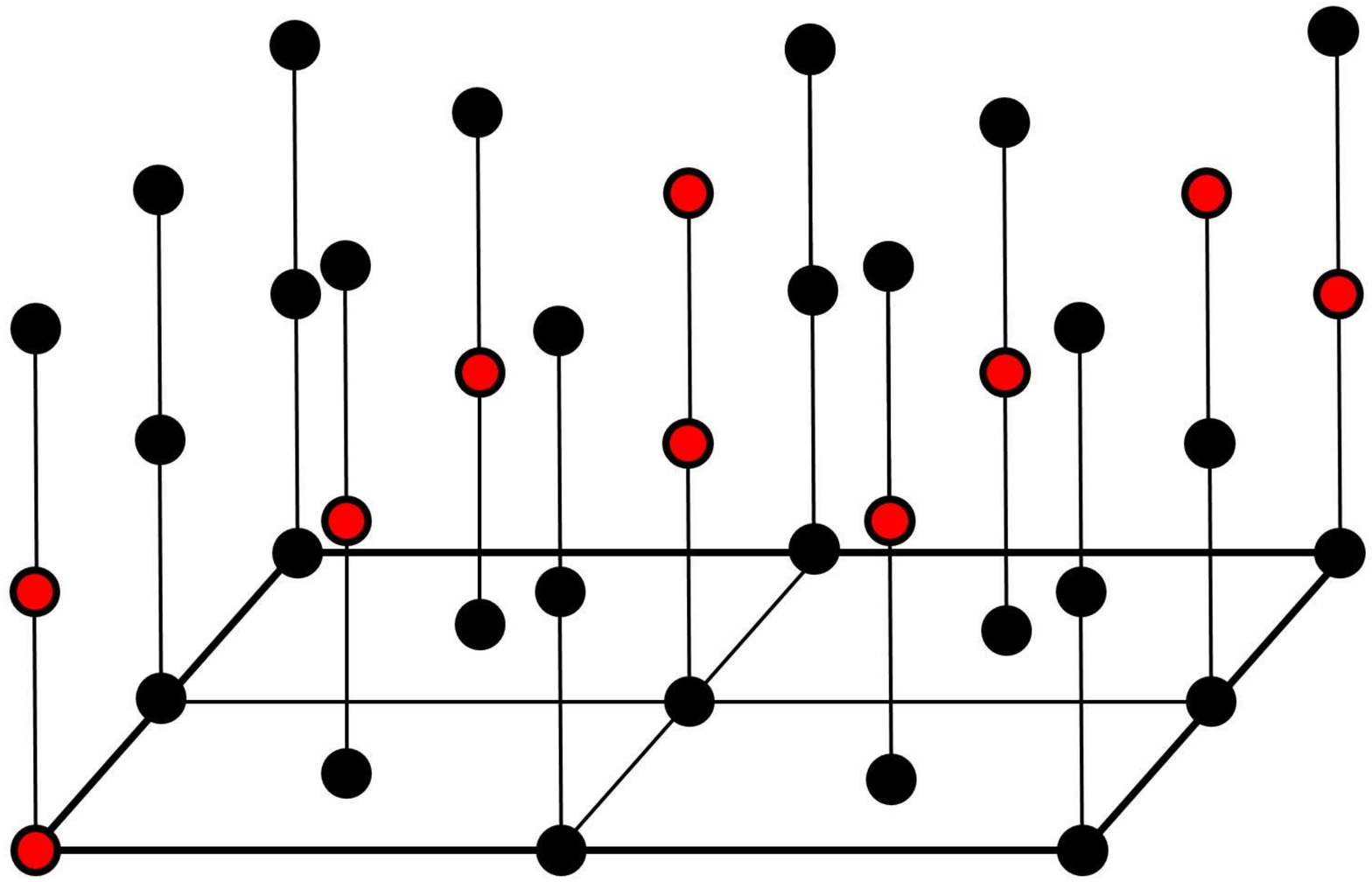}}
\hspace{0.2in}
\subfigure[]{
\label{fig9a}
\includegraphics[width=3cm]{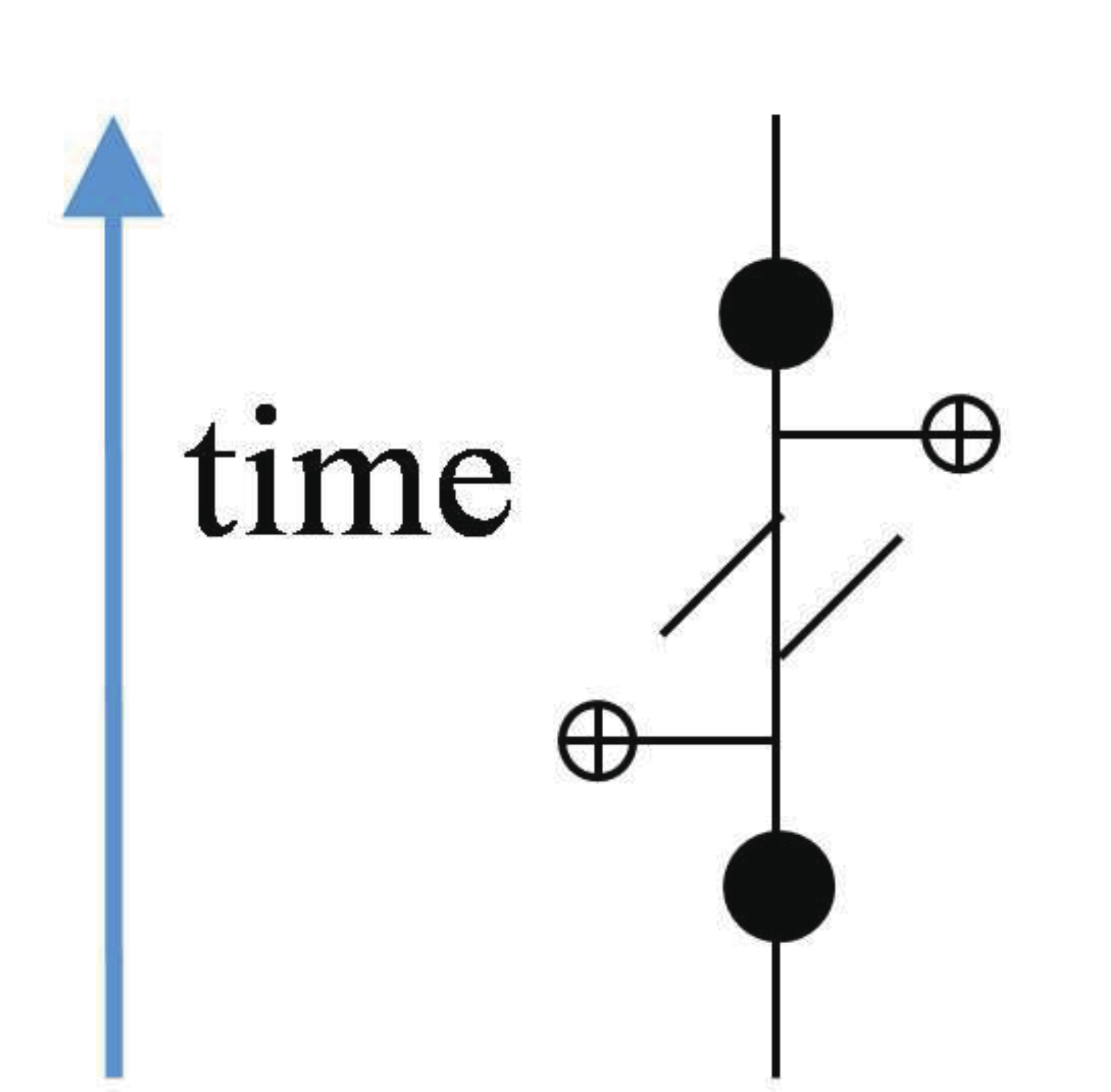}}
\hspace{0.4in}
\vspace{0.4in}
\subfigure[]{
\label{fig9b}
\includegraphics[width=4.5cm]{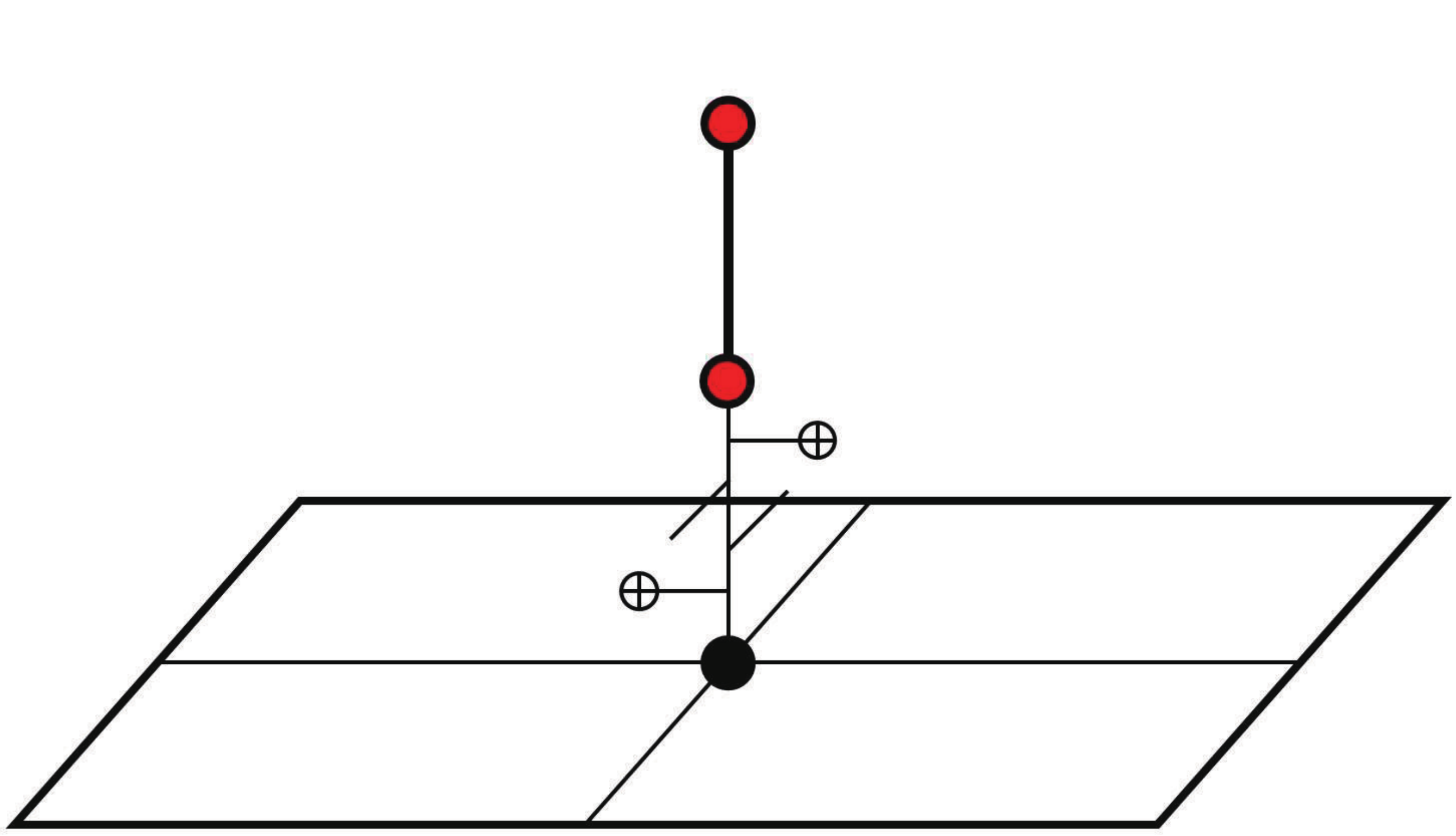}} 
\hspace{0.2in}
\subfigure[]{
\label{fig9c}
\includegraphics[width=4.5cm]{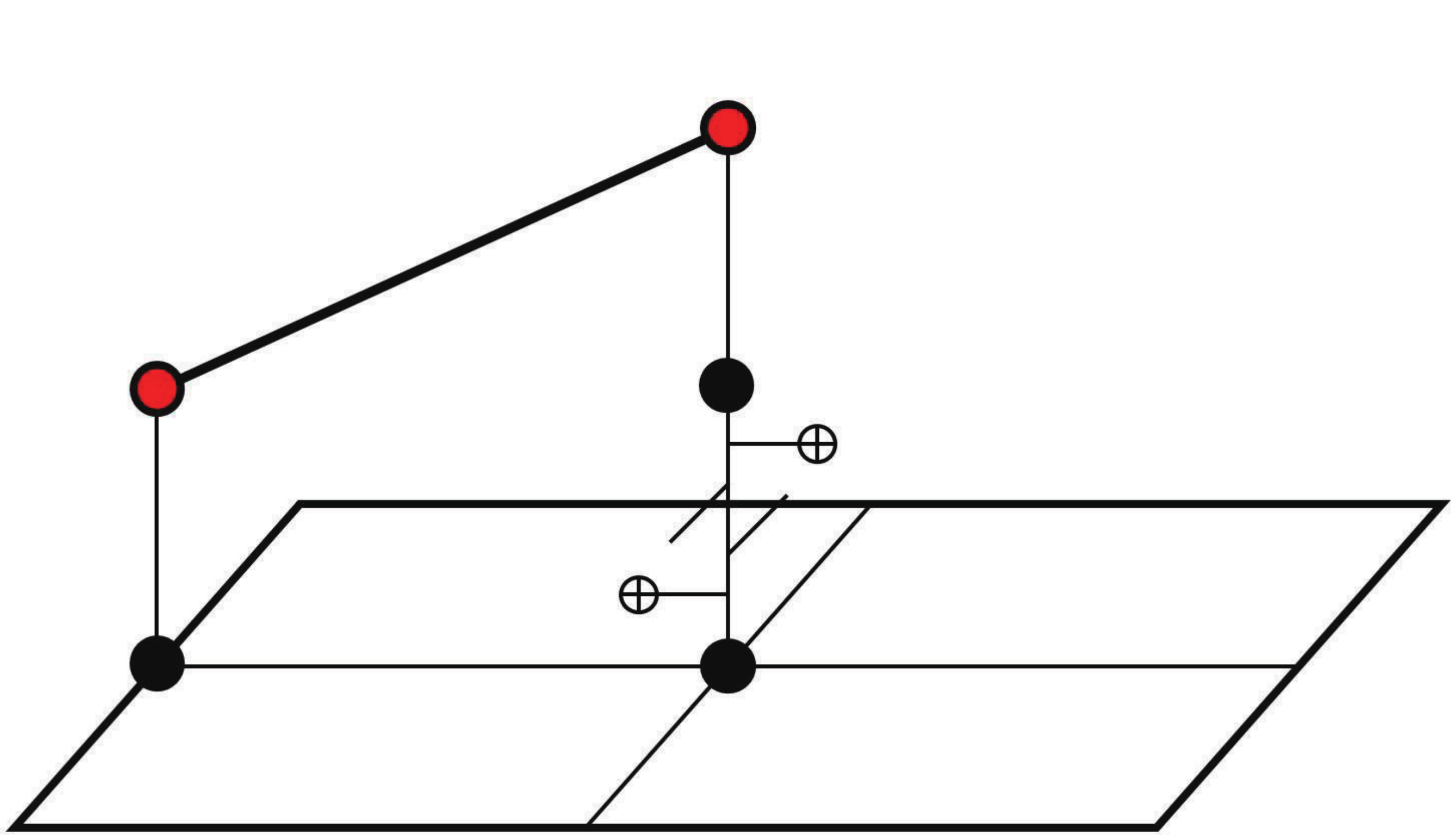}}
\hspace{0.2in} 
\subfigure[]{
\label{fig9d}
\includegraphics[width=4.5cm]{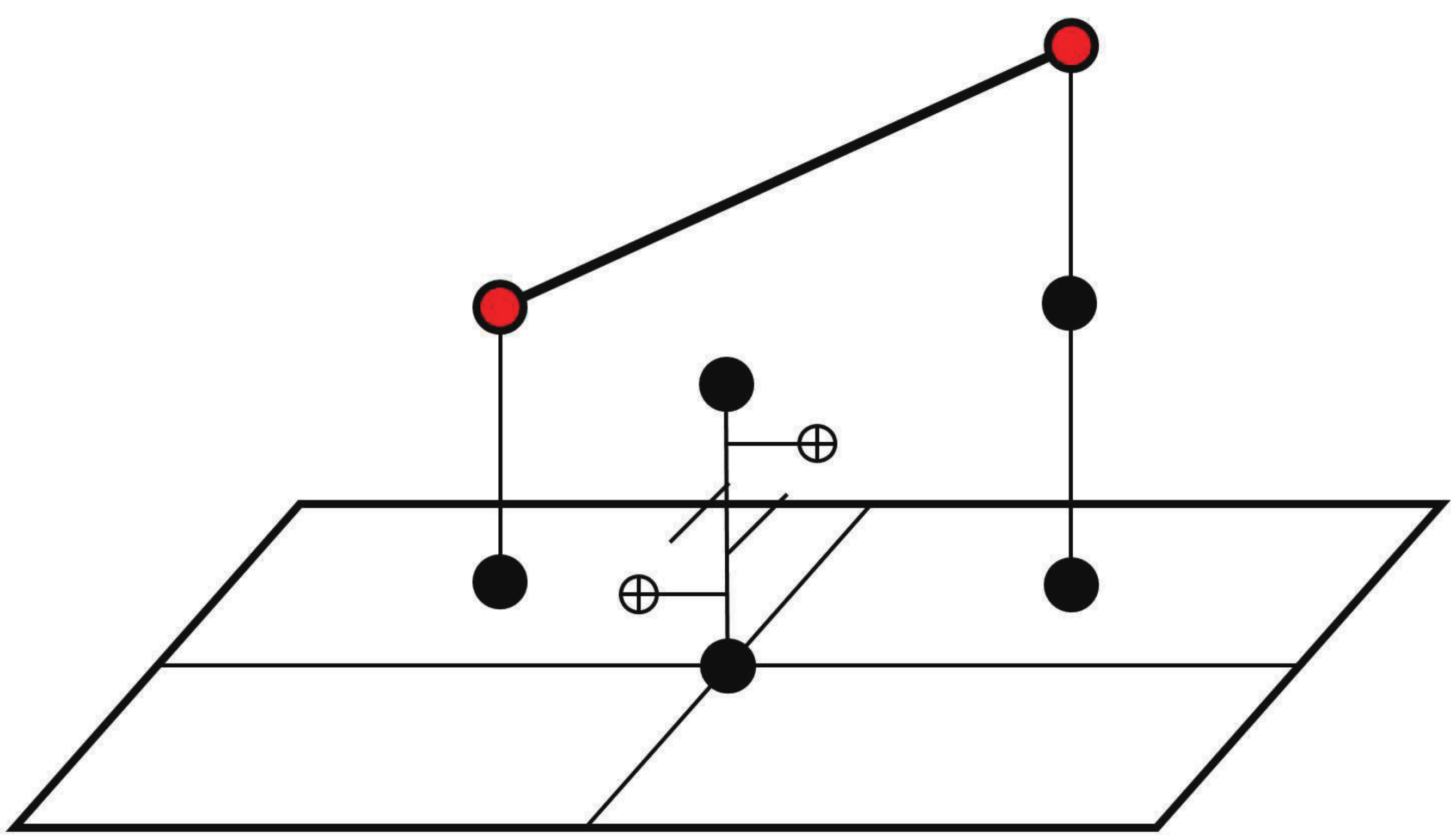}}
\hspace{0.2in}
\vspace{0.4in}
\subfigure[]{
\label{fig9e}
\includegraphics[width=4.5cm]{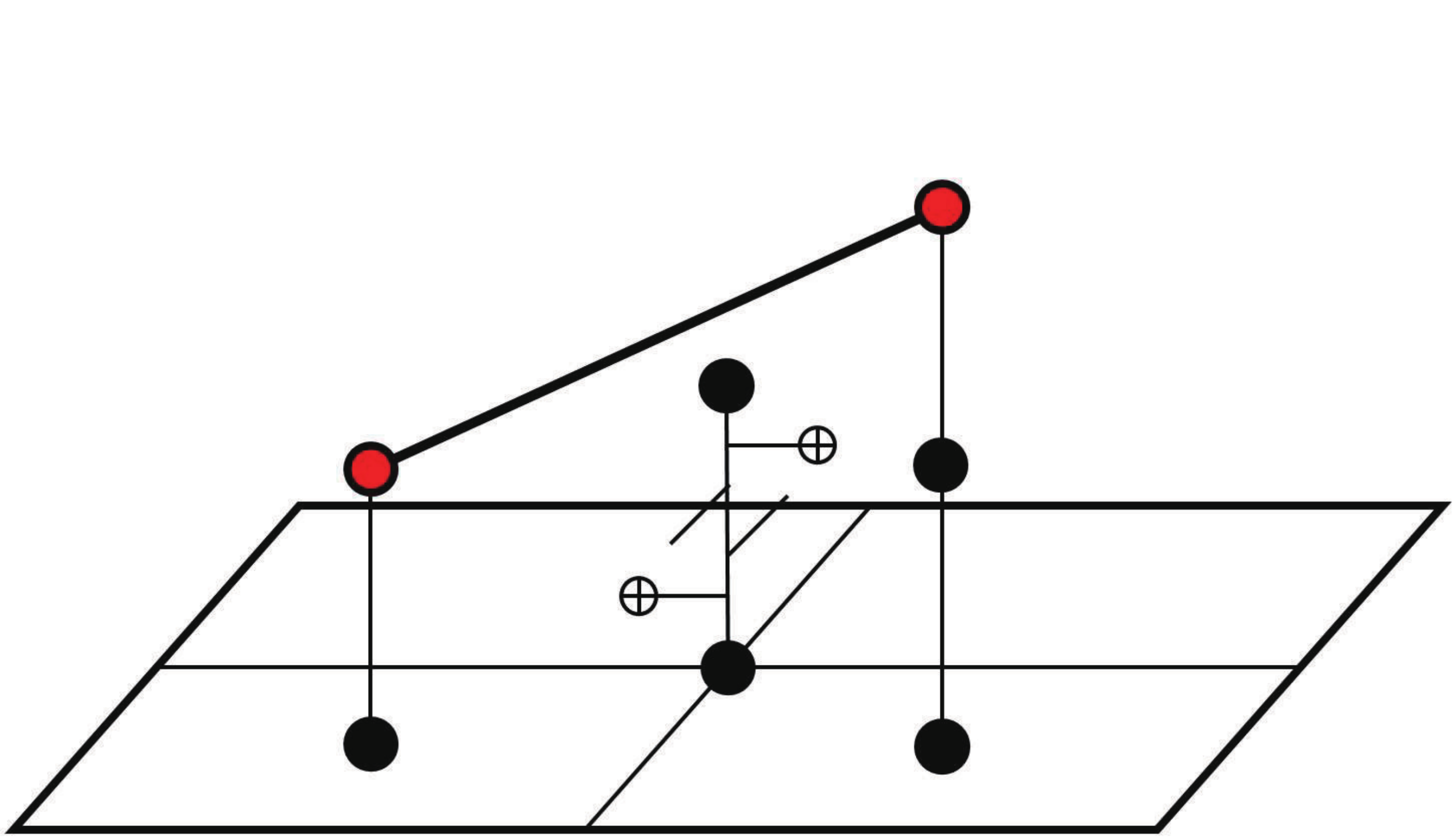}} 
\hspace{0.2in}
\subfigure[]{
\label{fig9f}
\includegraphics[width=4.5cm]{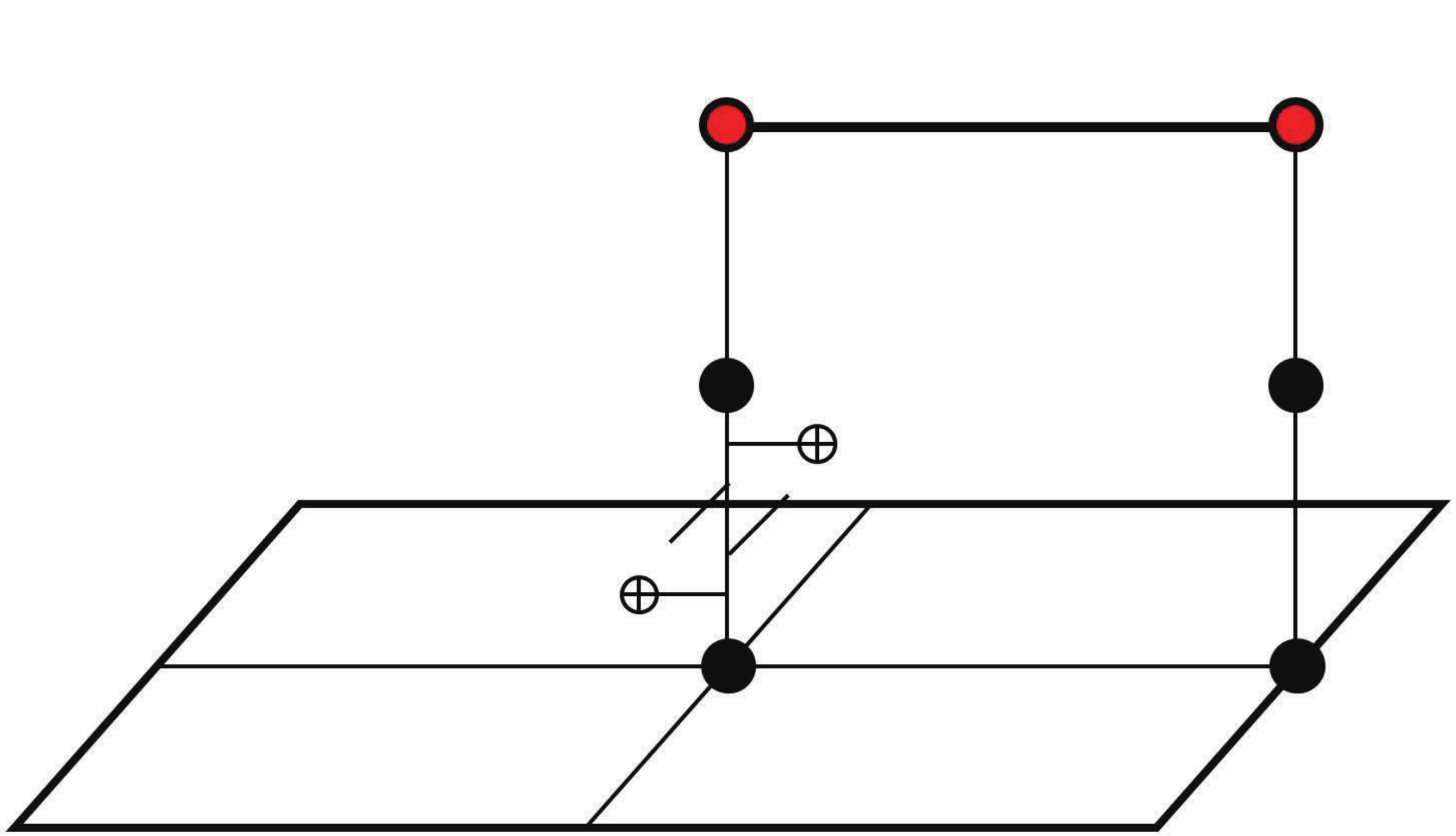}} 
\hspace{0.2in}
\subfigure[]{
\label{fig9g}
\includegraphics[width=4.5cm]{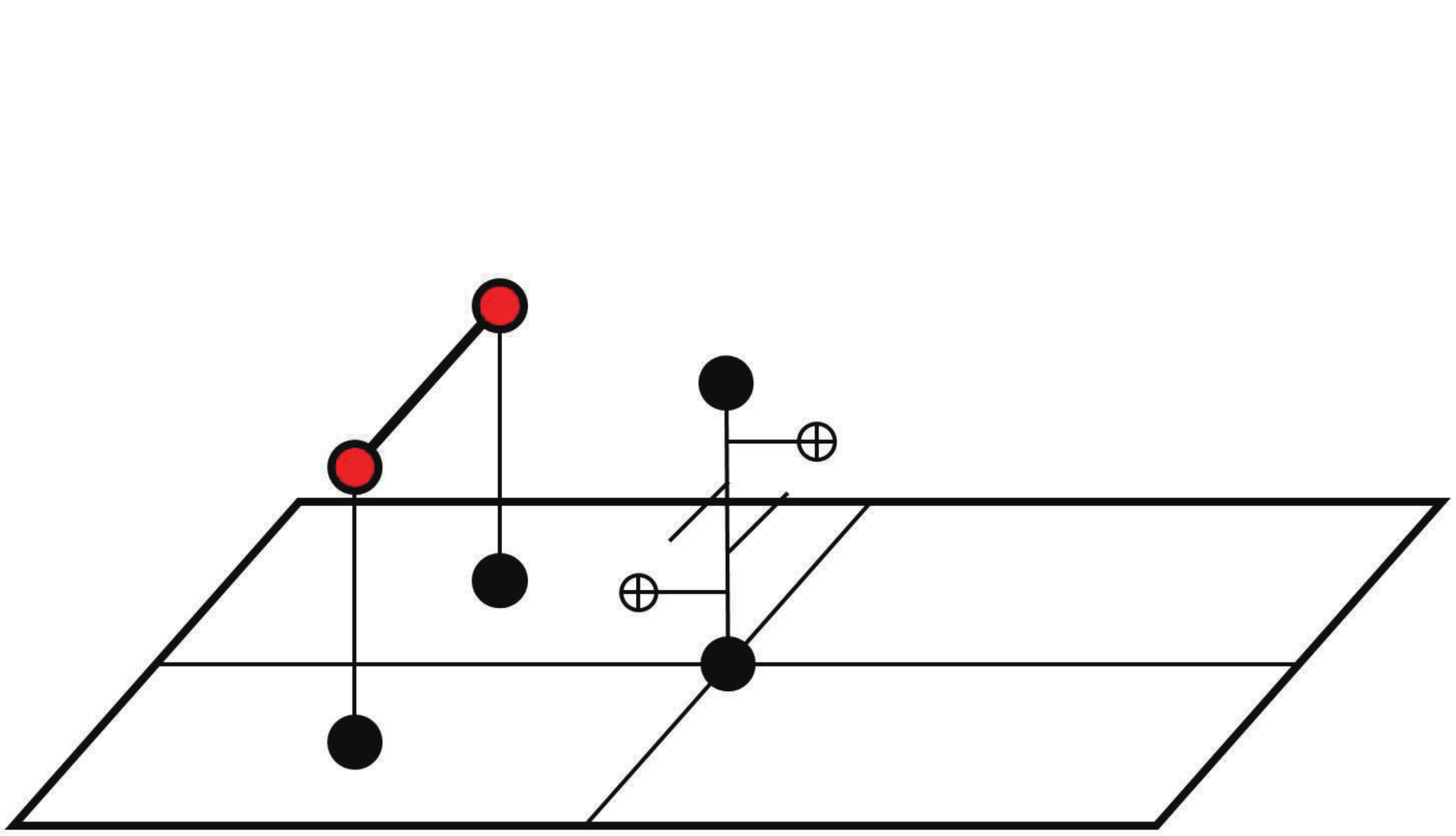}}
\hspace{0.2in}
\vspace{0.4in}
\subfigure[]{
\label{fig9h}
\includegraphics[width=4.5cm]{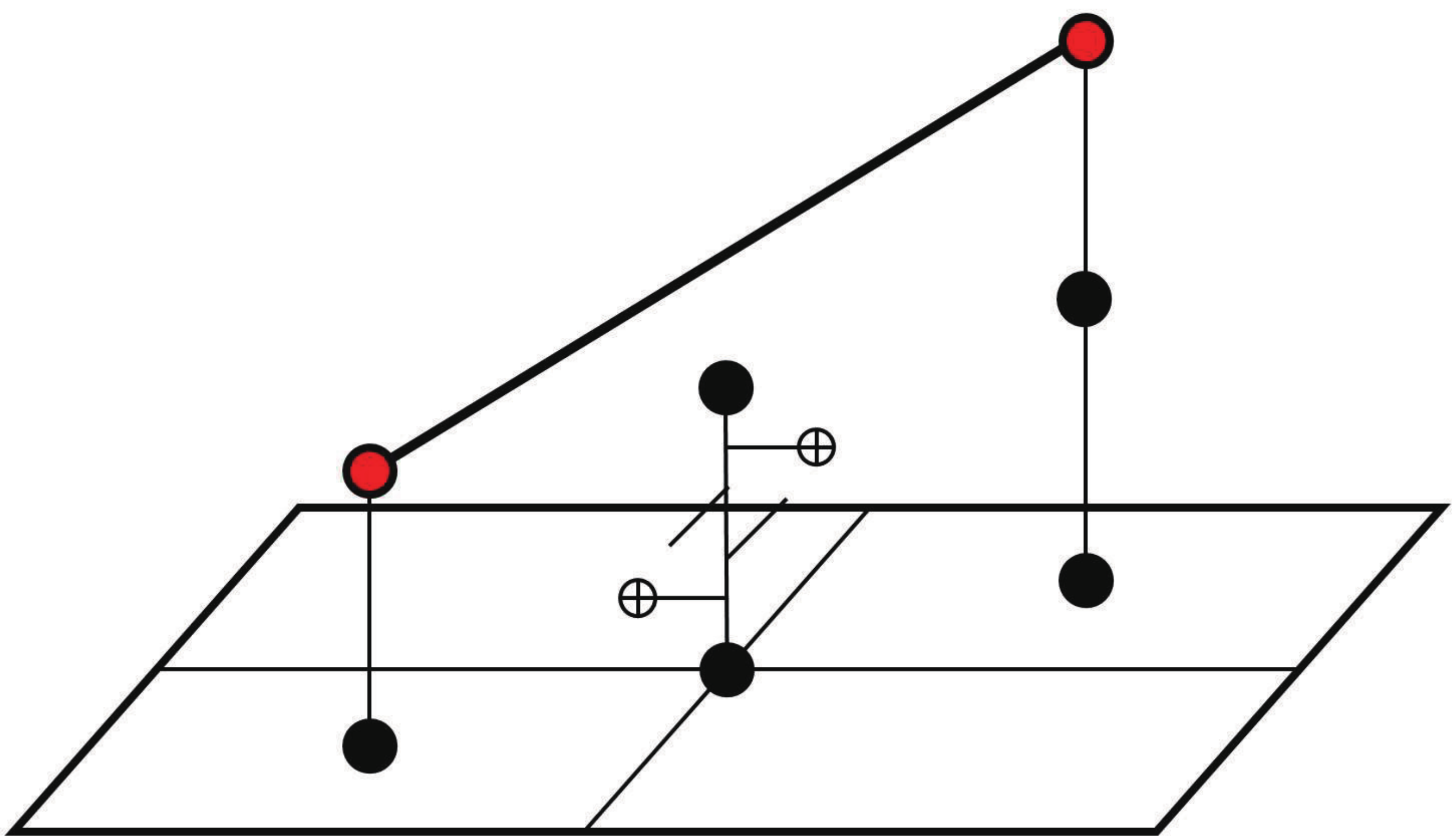}} 
\hspace{0.2in}
\subfigure[]{
\label{fig9i}
\includegraphics[width=4.5cm]{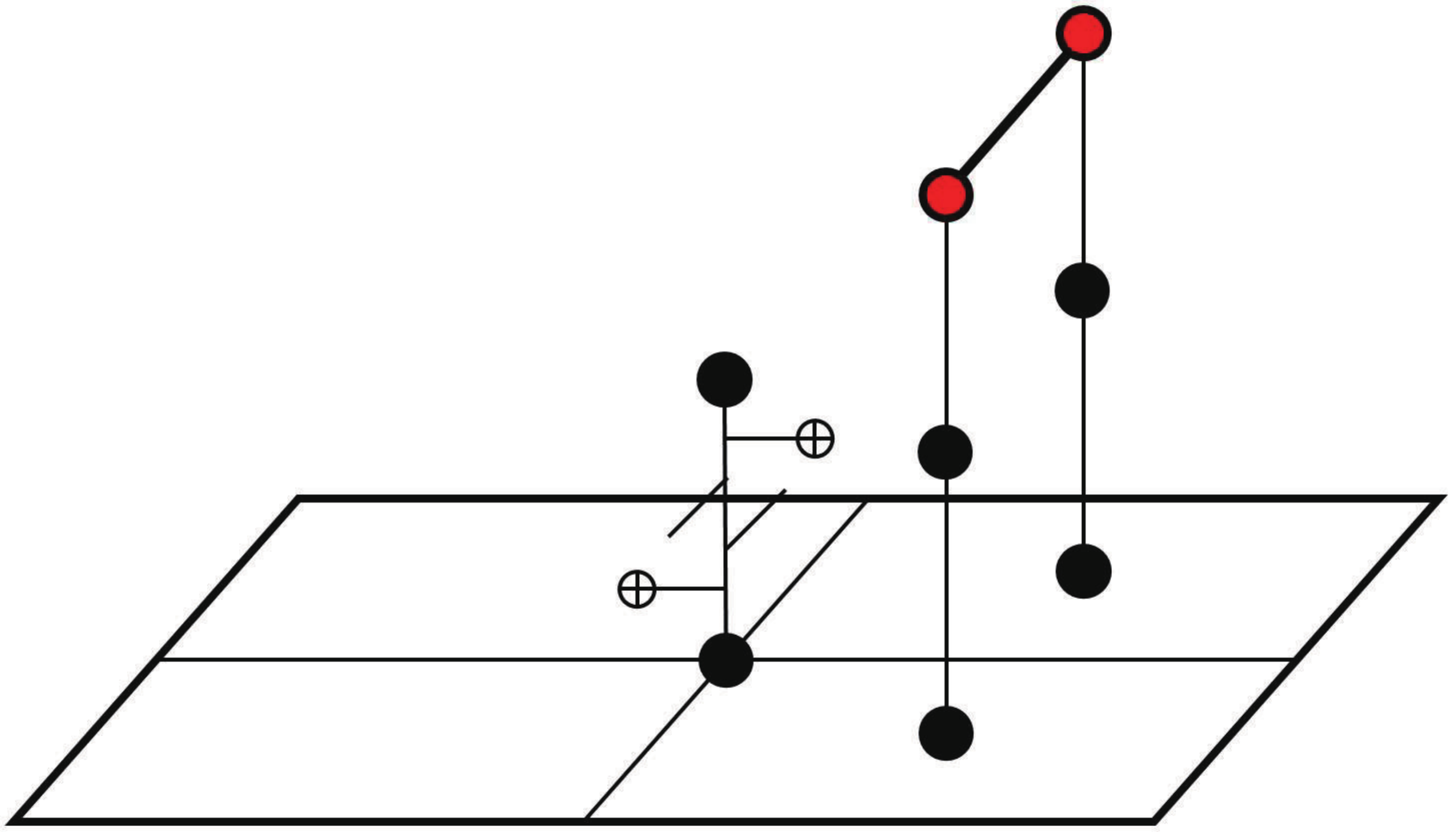}} 
\hspace{0.2in}
\subfigure[]{
\label{fig9j}
\includegraphics[width=4.5cm]{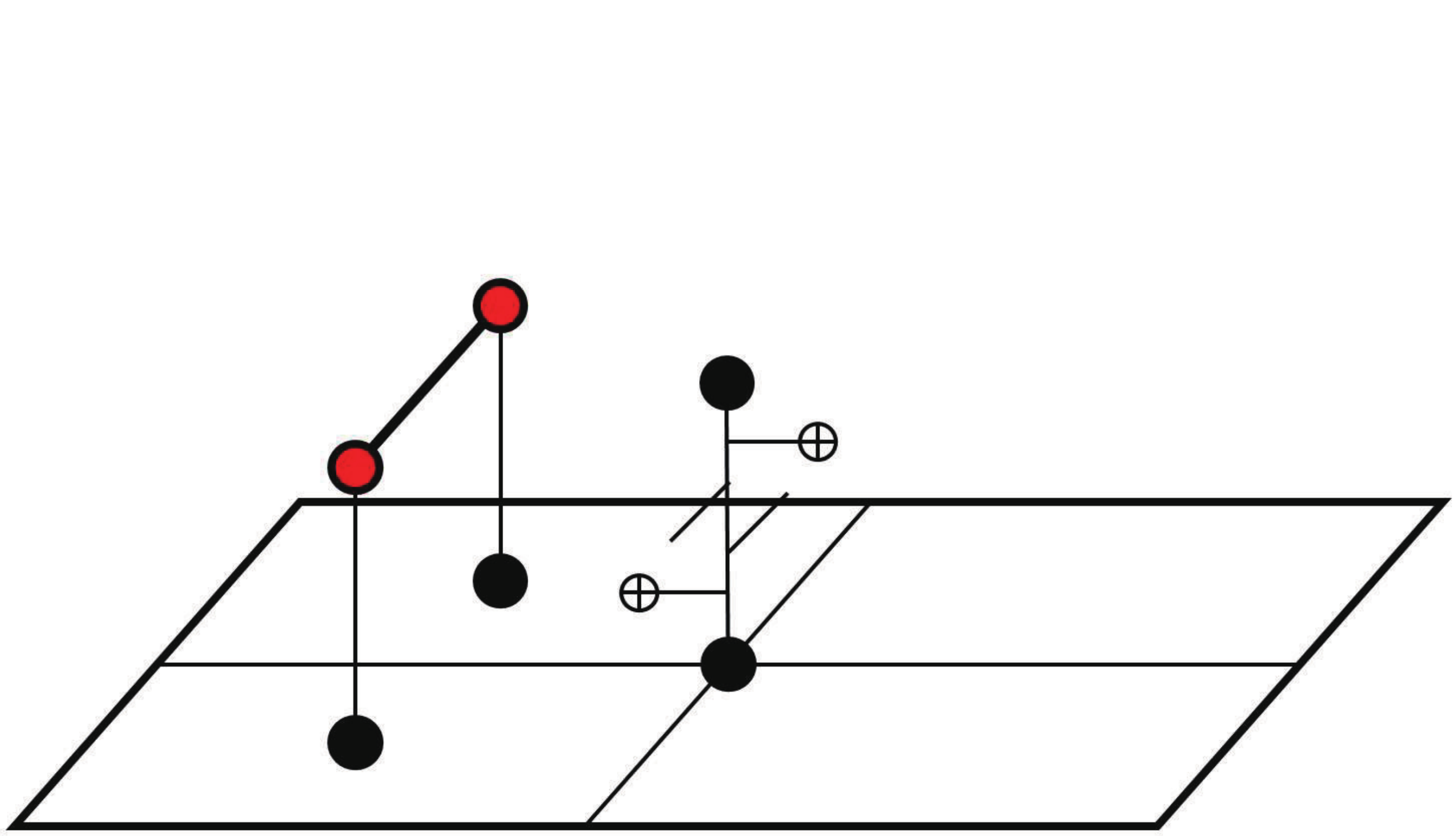}}
\hspace{0.2in}
\vspace{0.4in}
\subfigure[]{
\label{fig9k}
\includegraphics[width=4.5cm]{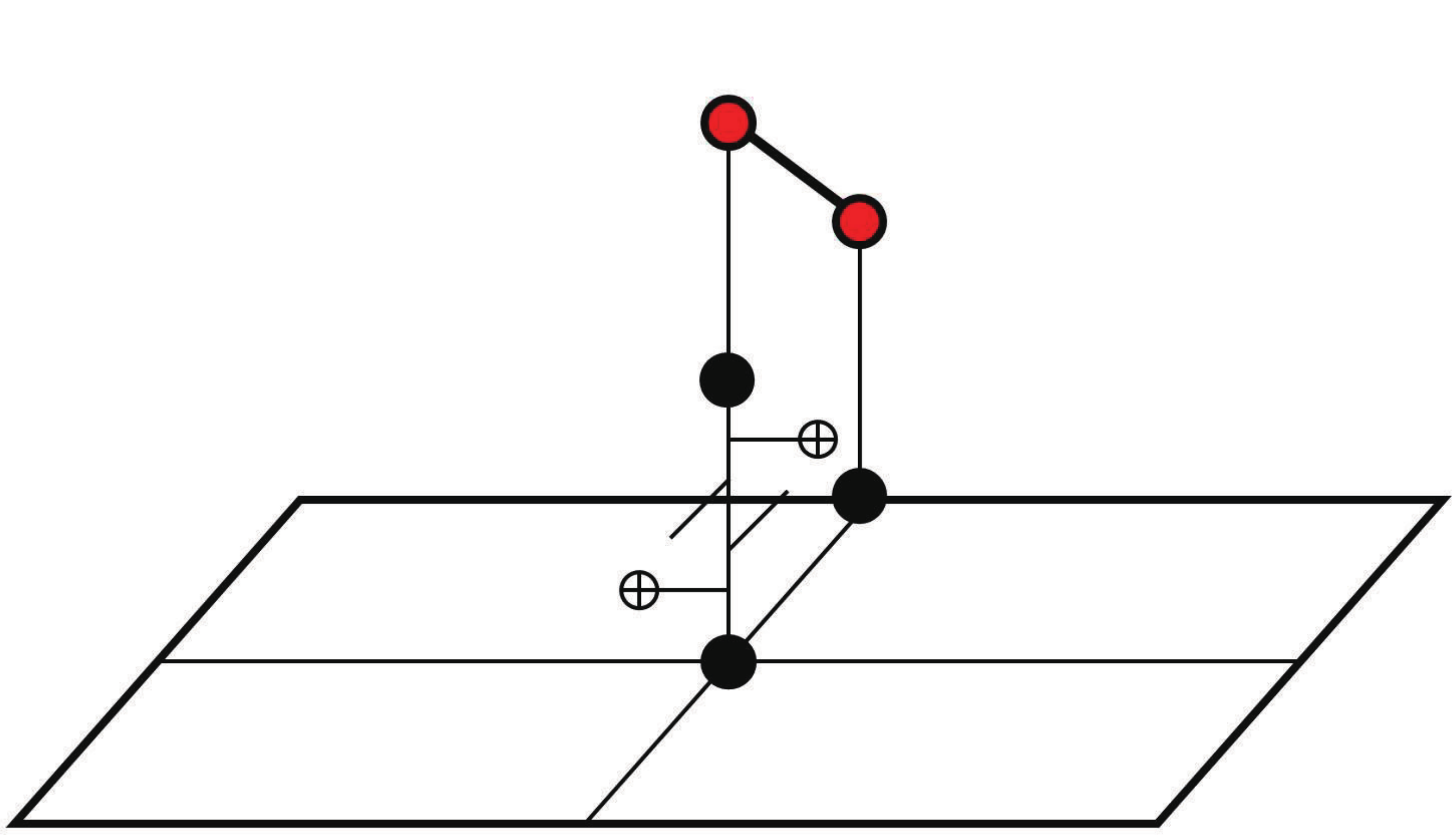}} 
\hspace{0.2in}
\subfigure[]{
\label{fig9l}
\includegraphics[width=4.5cm]{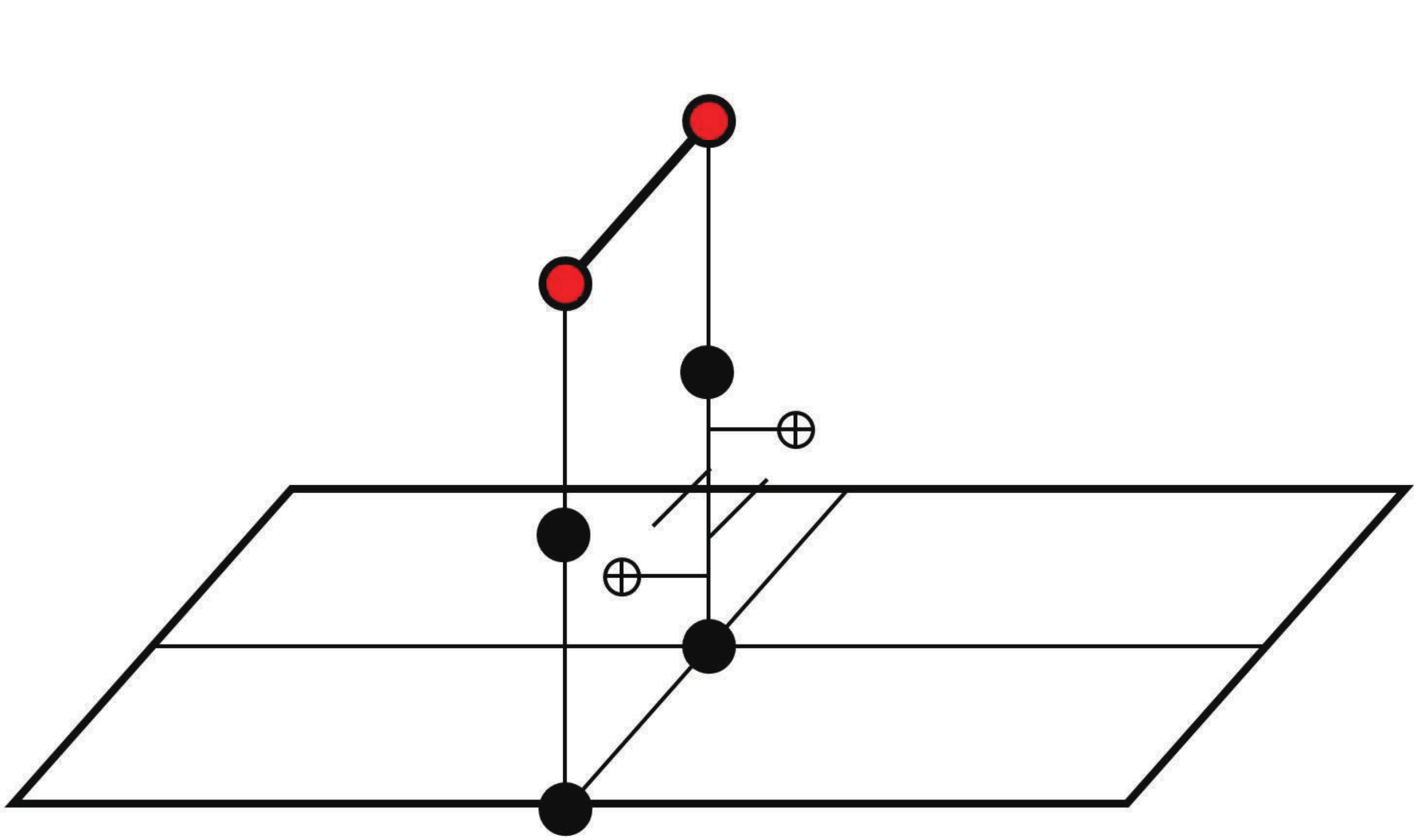}} 
\hspace{0.2in}
\subfigure[]{
\label{fig9m}
\includegraphics[width=4.5cm]{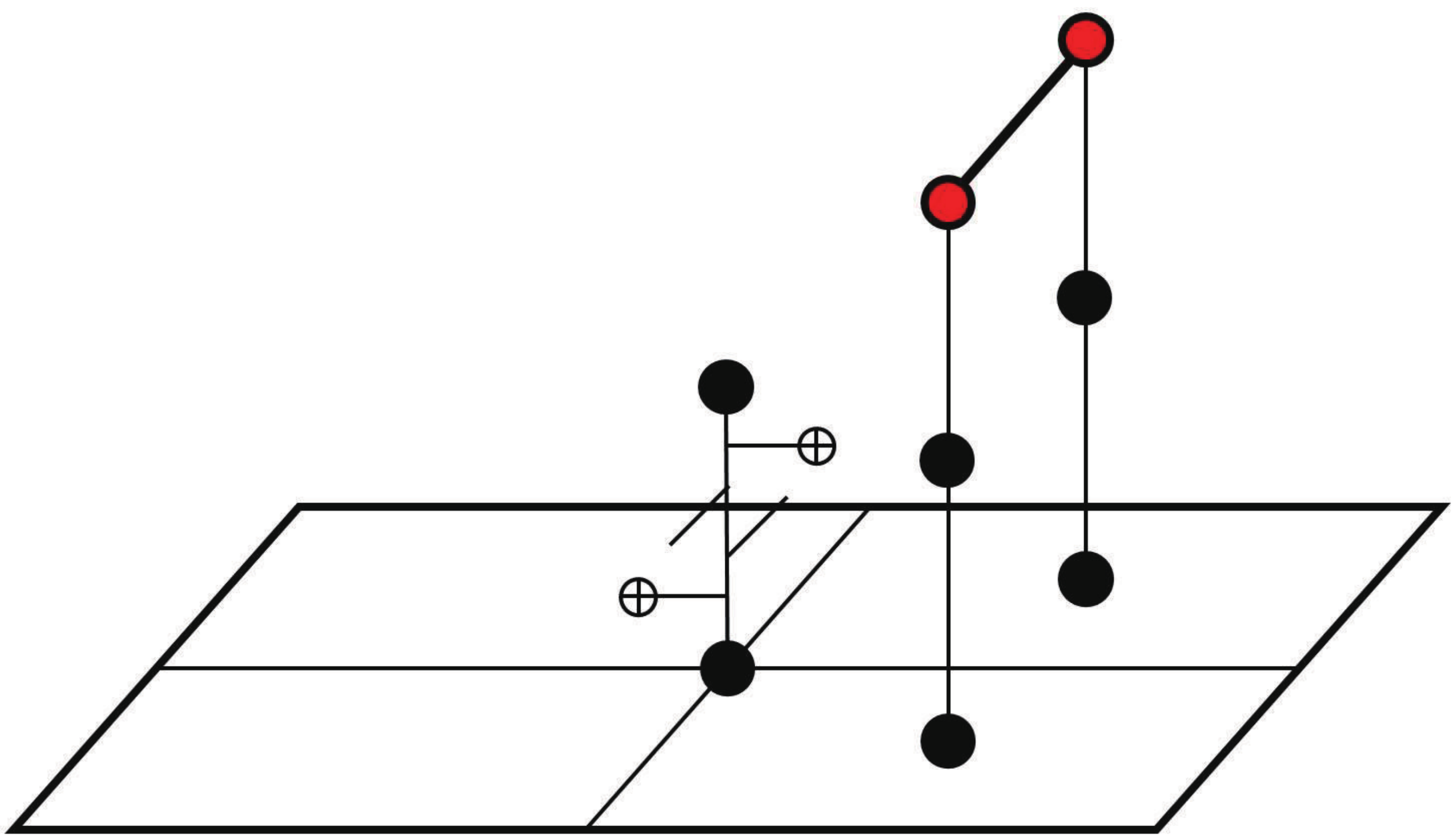}} 
\caption{\justify (a) Decoding graph of the XZZX surface code. The red vertices need to br matched in pair. (b) Four time steps of applying CNOT or CZ gates. Between syndrome qubit measurements, the four two-qubit gates are executed sequentially. (c) Syndromes of the $IZ$ error after 1-4 time steps. These syndromes can also be caused by the measurement errors of the syndrome qubits. (d-g) Syndromes of the $ZI$ error after 1-4 time steps respectively.  (h-k) Syndromes of the $IX$ error after 1-4 time steps respectively. (l-n) Syndromes of the $IX$ error after 1-3 time steps respectively. The syndromes of $IX$ error after 4th time step is trivial.}\label{fig9}
\end{figure*}

The GKP error correction provides continuous-variable information for the surface code decoding. Here we show the matching weights of the MWPM decoder when decoding the XZZX surface code, which is related to the conditional error rates of the GKP states. The decoding graph is constructed in a three-dimensional lattice as shown in Fig.$\,$\ref{fig90}, where each edge in the bottom face represents a data qubit and the labeled vertex (red) needs to be matched in pair. The stabilizer measurement circuit applies CNOT or CZ gates in four time step (see Fig.$\,$\ref{fig9a}). The error after each time step leads to a edge that connects two labeled vertex. The edges in Fig.$\,$\ref{fig9} can be divided into six types: \{c\}, \{d,e,f\}, \{g\}, \{h,j,k,m,n\}, \{i\}, \{l\}.  We assign the weights $w=-\log p$ to each type of edges, where $p$ is the sum of the probabilities of the errors that correspond to the same type of edges. Specifically speaking, the conditional probabilities of the $XI$ error and $IX$ error (the first operator acts on the data qubit and the second opertor acts on the syndrome qubit) after the CNOT gate are
\begin{equation}
\begin{aligned}
&p(XI|q_{m1},q_{m2})\\
=&\frac{\sum_{k_1,k_2\in\mathbb{Z}}p(u_1-(2k_1+1)\sqrt{\pi}\lambda,u_2-k_2\sqrt{\pi}\beta)}
{\sum_{k_1,k_2\in\mathbb{Z}}p(u_1-k_1\sqrt{\pi}\lambda,u_2-k_2\sqrt{\pi}\beta)},
\end{aligned}
\end{equation}

\begin{equation}
\begin{aligned}
&p(IX|q_{m1},q_{m2})\\
=&\frac{\sum_{k_1,k_2\in\mathbb{Z}}p(u_1-k_1\sqrt{\pi}\lambda,u_2-(2k_2+1)\sqrt{\pi}\beta)}
{\sum_{k_1,k_2\in\mathbb{Z}}p(u_1-k_1\sqrt{\pi}\lambda,u_2-k_2\sqrt{\pi}\beta)},
\end{aligned}
\end{equation}
where $p(x,y)$ is the probability density function in Eq.$\,$(\ref{e32}) and $u_1=\sqrt{2}q_{m1}-n_1\sqrt{\pi}\lambda$, $u_2=\sqrt{2}q_{m2}-n_2\sqrt{\pi}\lambda$. Note that $q_{m1}$, $q_{m2}$ are the measurement results in $\hat{q}$ quadrature and $n_1$, $n_2$ are obtained by the ML decoding in \ref{4b}. Since the probability of $XX$ error usually is much smaller than that of $XI$ or $IX$ error, we do not consider the edge caused by $XX$ error independently, but view it as the superposition of $XI$ and $IX$.  

Likewise, it is easy to compute the the conditional probability of the $ZI$ or $IZ$ errors after the CNOT or CZ gate by substituting the measurement results and the probability density functions in $\hat{p}$ quadrature. 

Moreover, we determine the syndrome as +1 if $|p_m\mod{2\sqrt{\pi}}/\beta|<\frac{\sqrt{\pi}}{2\beta}$, othewise it is -1, where $p_m$ is the measurement result of the syndrome qubit. The error probability of the measurement is calculated by: 
\begin{equation}
\begin{aligned}
p_{meas}=1-\frac{\sum_{k\in\mathbb{Z}}P_\sigma (p_m-2k\sqrt{\pi}/\beta)}{\sum_{k\in\mathbb{Z}}P_\sigma (p_m-k\sqrt{\pi}/\beta)},
\end{aligned}
\end{equation}
where $\sigma=\sqrt{\sigma_p^2+\sigma_c^2+\sigma_m^2}$.

Lastly, when syndrome qubits are measured, data qubits are applied GKP error corrections with the error probabilities
\begin{equation}
\begin{aligned}
p_{idle}(X)=1-\frac{\sum_{k\in\mathbb{Z}}P_\sigma' (q_m-2k\sqrt{\pi}\lambda)}{\sum_{k\in\mathbb{Z}}P_\sigma' (q_m-k\sqrt{\pi}\lambda)},
\end{aligned}
\end{equation}
\begin{equation}
\begin{aligned}
p_{idle}(Z)=1-\frac{\sum_{k\in\mathbb{Z}}P_\sigma' (p_m-2k\sqrt{\pi}/\lambda)}{\sum_{k\in\mathbb{Z}}P_\sigma' (p_m-k\sqrt{\pi}/\lambda)},
\end{aligned}
\end{equation}
where $\sigma'=\sqrt{\sigma_p^2+2\sigma_c^2+\sigma_m^2+\sigma_i^2}$. 

\newpage
\bibliographystyle{unsrt}
\bibliography{gkp}

\end{document}